\documentclass[fleqn,usenatbib]{mnras}

\usepackage{newtxtext,newtxmath}

\usepackage[T1]{fontenc}

\DeclareRobustCommand{\VAN}[3]{#2}
\let\VANthebibliography\thebibliography
\def\thebibliography{\DeclareRobustCommand{\VAN}[3]{##3}\VANthebibliography}

\usepackage[version=4]{mhchem}
\usepackage{chemmacros}
\chemsetup{
  modules = {reactions} ,
  formula = mhchem
}

\chemsetup[reactions]{
  before-tag = R ,
  tag-open = ( ,
  tag-close = )
}
\usepackage{graphicx}	
\usepackage{amsmath}	

\title[Circulation and ozone on synchronous exoplanets]{Stratospheric dayside-to-nightside circulation drives the 3-D ozone distribution on synchronously rotating rocky exoplanets}
\author[M. Braam et al.]{
Marrick Braam,$^{1,2,3}$\thanks{E-mail: \href{mailto:mbraam@ed.ac.uk}{mbraam@ed.ac.uk}}
Paul I. Palmer$^{1,2}$,
Leen Decin$^{3}$,
Maureen Cohen$^{1,2}$,
and Nathan J. Mayne$^{4}$\\
$^{1}$School of GeoSciences, University of Edinburgh, Edinburgh, EH9 3FF, UK\\
$^{2}$Centre for Exoplanet Science, University of Edinburgh, Edinburgh, EH9 3FD, UK\\
$^{3}$Institute of Astronomy, KU Leuven, 3001 Leuven, Belgium\\
$^{4}$Department of Physics and Astronomy, Faculty of Environment Science and Economy, University of Exeter, Exeter, EX4 4QL, UK\\
}

\date{Accepted XXX. Received YYY; in original form ZZZ}

\pubyear{2022}

\begin{document}
\label{firstpage}
\pagerange{\pageref{firstpage}--\pageref{lastpage}}
\maketitle

\begin{abstract}
Determining the habitability and interpreting future atmospheric observations of exoplanets requires understanding the atmospheric dynamics and chemistry from a 3-D perspective. Previous studies have shown significant spatial variability in the ozone layer of synchronously rotating M-dwarf planets, assuming an Earth-like initial atmospheric composition. We use a 3-D Coupled Climate-Chemistry model to understand this distribution of ozone and identify the mechanism responsible for it. We document a previously unreported connection between the ozone production regions on the photochemically active dayside hemisphere and the nightside devoid of stellar radiation and thus photochemistry. We find that stratospheric dayside-to-nightside overturning circulation can advect ozone-rich air to the nightside. On the nightside, ozone-rich air subsides at the locations of two quasi-stationary Rossby gyres, resulting in an exchange between the stratosphere and troposphere and the accumulation of ozone at the gyre locations. We identify the hemispheric contrast in radiative heating and cooling as the main driver of this ozone circulation. Dynamically-driven chemistry also impacts other tracer species in the atmosphere (gaseous and non-gaseous phase) as long as chemical lifetimes exceed dynamical lifetimes. These findings illustrate the 3-D nature of planetary atmospheres, predicting spatial and temporal variability that will impact spectroscopic observations of exoplanet atmospheres.
\end{abstract}

\begin{keywords}
Planets and satellites: terrestrial planets -- Planets and satellites: atmospheres -- Planets and satellites: composition
\end{keywords}



\section{Introduction}\label{sec:intro}
The past two decades have seen the discovery of numerous Earth-size exoplanets, with a substantial fraction of them orbiting in the circumstellar Habitable Zone \citep[][]{kasting_habitable_1993}. Earth-size planets are preferentially discovered around M-dwarf stars \citep[][]{dressing_occurrence_2015}, because they are the most abundant stellar type, have relatively small radii, and are relatively cool, allowing for exoplanets in short-period orbits. The habitability of such exoplanets has been debated in light of the stellar and planetary environments \citep[][]{shields_habitability_2016}. Comprehensive numerical simulations that describe the physical and chemical properties of a planetary atmosphere in such environments are essential to understanding habitability and interpreting spectroscopic observations.

Since M stars are cooler and smaller than other stellar types, a planet in the Habitable Zone orbits at a small orbital distance and feels a strong gravitational pull from the host star. This can lead to spin-orbit resonances for the planet, so-called tidal locking, of which the most extreme case is the 1:1 resonant orbit or synchronous rotation \citep[e.g.][]{barnes_tidal_2017, renaud_tidal_2021}. Simulations with General Circulation Models (GCMs) help us understand how synchronous rotation affects the planetary atmosphere and surface habitability. First, synchronous rotation creates distinct hemispheric environments and a large temperature difference between the dayside and nightside \citep[e.g.][]{joshi_simulations_1997}. Second, synchronous rotation leads to distinct photochemical environments, with strong photochemical production and destruction on the dayside and an absence of photochemistry on the nightside \citep[e.g.][]{proedrou_characterising_2016, chen_biosignature_2018, yates_ozone_2020, braam_lightning-induced_2022, ridgway_3d_2023}. Depending on the rotation period, synchronous rotation can also lead to atmospheric circulation that is characterised by thermally direct circulation for slowly rotating planets \citep[e.g.][]{merlis_atmospheric_2010, edson_atmospheric_2011, heng_atmospheric_2011, koll_temperature_2016, haqq-misra_demarcating_2018}. The existence of this large-scale circulation requires the Rossby deformation radius to exceed the planetary radius \citep[][]{carone_connecting_2014, carone_connecting_2015, noda_circulation_2017, haqq-misra_demarcating_2018}, which is the case for planets like Proxima Centauri b, Trappist-1 e to h, LHS-1140 b and GJ 667 C c, assuming an Earth-like atmosphere. The dayside-nightside contrast leads to an overturning circulation, with upwelling on the dayside and downwelling on the nightside \citep[][]{showman_atmospheric_2013}. This vertical motion results in a superposition of planetary-scale Rossby and Kelvin waves, which drives eddy momentum equatorward \citep[][]{showman_matsuno-gill_2010}. A typical part of this wave structure is a pair of quasi-stationary cyclonic gyres on the nightside \citep[][]{showman_matsuno-gill_2010}. The equatorward momentum feeds the superrotating jet \citep[][]{showman_equatorial_2011}. The overturning circulation is a dominant component of the dayside-to-nightside heat transport \citep[][]{hammond_rotational_2021}. 

Atmospheric circulation impacts the spatial and temporal distribution of chemical species and other tracers such as clouds \citep[e.g.][]{boutle_exploring_2017, komacek_atmospheric_2019, sergeev_atmospheric_2020} and photochemical hazes \citep[][]{parmentier_3d_2013, steinrueck_3d_2021}.  On Earth, the Brewer-Dobson circulation controls the large-scale distribution of chemical tracers such as ozone (O$_3$) and water vapour in the atmosphere \citep[][]{butchart_brewer-dobson_2014}. Ozone formation is initiated by photochemistry through the Chapman mechanism \citep[][]{chapman_xxxv_1930}, which is strongest at tropical latitudes. The Brewer-Dobson circulation describes the ascent of ozone-rich air in the tropics, followed by equator-to-pole transport and descending air motions at high latitudes, leading to meridional variations with a relatively enhanced ozone layer at high latitudes. 


\citet{proedrou_characterising_2016} simulated a tidally-locked Earth using a 3-D climate-chemistry model (CCM), which consists of a GCM coupled to a photochemical network to study the relation between (photo)chemistry, atmospheric dynamics and the thermal structure of the atmosphere. They find a breakdown of the Brewer-Dobson circulation, and instead predict that ozone accumulates on the nightside, where it has a long lifetime \citep[][]{proedrou_characterising_2016}. \citet{carone_stratosphere_2018} investigated stratospheric circulation on tidally-locked exoplanets and the potential impact on the distribution of chemical species. For planets with short orbital periods (${<}25$~days), tropical Rossby waves can induce strong equatorial jets in the stratosphere with pole-to-equator transport of airmasses \citep[][]{carone_stratosphere_2018}. \citet{chen_habitability_2019} showed the meridional distribution of ozone from CCM simulations, confirming that this pole-to-equator circulation essentially confines photochemical species such as ozone to the equatorial regions. The existence of extratropical Rossby waves or damping of tropical Rossby waves prevents this equatorial confinement. Instead, a thermally-driven overturning circulation can drive equator-to-pole transport of photochemical species \citep[][]{carone_stratosphere_2018, chen_habitability_2019}, leading to meridional structure with enhanced ozone at high latitudes. For planets like Proxima Centauri b, \citet{carone_stratosphere_2018} find a relatively weak tropical Rossby wave, with a thermally-driven equator-to-pole circulation existing in the stratosphere (see their Figure 12). For such planets, the enhanced ozone abundances at high latitudes were later also simulated by \citet{chen_habitability_2019}.

The distribution of radiatively active species such as ozone impacts habitability \citep[e.g.][]{ridgway_3d_2023}, and will determine what spectroscopic observations of the planetary atmosphere will look like \citep[e.g.][]{cooke_variability_2023, zamyatina_observability_2023}. Despite reporting a non-detection for the atmosphere, the observation of TRAPPIST-1 b illustrates the capability of JWST to characterise Earth-size exoplanets \citep[][]{greene_thermal_2023}. For the exoplanets that have an atmosphere we need to understand their 3-D nature, including circulation, clouds, and atmospheric chemistry, which motivates the application of 3-D CCMs to exoplanetary environments. Such simulations of synchronously rotating exoplanets predict a significant zonal structure in the ozone layer for planets around M-dwarfs like Proxima Centauri b \citep[][]{yates_ozone_2020, braam_lightning-induced_2022} and haze distribution for hot Jupiters \citep[][]{parmentier_3d_2013, steinrueck_3d_2021}. \citet{yates_ozone_2020} found that ozone has a much longer chemical lifetime on the nightside as compared to the dayside of M-dwarf exoplanets. These long nightside lifetimes lead to accumulation of ozone in the nightside gyres, despite the absence of stellar radiation needed to initiate the relevant photochemistry. This spatially variable ozone layer indicates a connection between the photochemically active dayside regions and nightside gyres, which is currently not understood.

In this paper, we aim to understand the dayside-nightside connection and identify the physical and chemical mechanism that drives the spatially variable ozone layer on synchronously rotating exoplanets around M-dwarf stars. We use a 3-D CCM to investigate the spatial and temporal structure of atmospheric ozone, using a configuration for Proxima Centauri b. In Section~\ref{sec:methods}, we briefly describe the CCM and introduce metrics used to diagnose atmospheric circulation. This will be followed by a description of the ozone distribution and its relation to atmospheric circulation in Section~\ref{sec:results}. In Section~\ref{sec:discussion}, we identify a possible driver of the circulation, investigate variability in our simulations and investigate potential observability. Finally, we present the conclusions of our study in Section~\ref{sec:conclusion}.

\section{Methods}\label{sec:methods}
This section starts with a description of the 3-D coupled climate-chemistry model. This is followed by the introduction of useful metrics to diagnose the atmospheric circulation and its impact on chemistry in Section~\ref{sec:metrics}. Finally, we summarize the experimental setup in Section~\ref{sec:setup}.
\subsection{Coupled Climate-Chemistry Model}
\label{sec:3dccm} 
The 3-D CCM consists of the Met Office Unified Model (UM) as the GCM coupled with the UK Chemistry and Aerosol framework (UKCA), in the configuration described by \citet{braam_lightning-induced_2022}. UM-UKCA is used to simulate the atmospheric dynamics and chemistry for Proxima Centauri b, but the results apply to other planets in similar orbits around M-dwarf stars. We simulate an aquaplanet with 1~bar or 1000 hPa surface pressure \citep[see][and references therein]{braam_lightning-induced_2022} and use a horizontal resolution of $2\degr$ by $2.5\degr$ in latitude and longitude, respectively. The atmosphere extends up to 85~km in 60 vertical levels. We assume that Proxima Centauri b is in a 1:1 resonant orbit around its M-dwarf host star and use the orbital parameters as shown in Table~\ref{tab:orbplan_params}. The substellar point is located at 0$^\circ$ latitude ($\phi$) and 0$^\circ$ longitude ($\lambda$). 

\begin{table}
	\centering
	\caption{Orbital and planetary parameters for the Proxima Centauri b setup, following \citet{boutle_exploring_2017}.}
	\label{tab:orbplan_params}
	\begin{tabular}{ll} 
		\hline
		Parameter & Value \\ 
            \hline
		Semi-major axis (AU) & 0.0485 \\
		Stellar Irradiance (W~m$^{-2}$) & 881.7 \\
		Orbital Period (days) & 11.186 \\
		Rotation rate (rad~s$^{-1}$) & $6.501\times10^{-6}$ \\
		Eccentricity & 0 \\
		Obliquity & 0 \\
		Radius (R$_\oplus$) & 1.1 \\
		Surface gravity (m~s$^{-2}$) & 10.9 \\
		\hline
	\end{tabular}
\end{table}

The UM is used in the Global Atmosphere 7.0 configuration \citep{walters_met_2019}, including the ENDGame dynamical core to solve the non-hydrostatic fully compressible deep-atmosphere equations of motion \citep[][]{wood_inherently_2014}. Parametrized sub-grid processes include convection (mass-flux approach, based on \citealt{gregory_mass_1990}), water cloud physics \citep[][]{wilson_pc2_2008}, turbulent mixing \citep[][]{lock_new_2000, brown_upgrades_2008} and the generation of lightning \citep[][]{price_simple_1992, luhar_assessing_2021, braam_lightning-induced_2022}. The incoming stellar radiation for 0.5~nm to 5.5~$\mu$m is described by the v2.2 composite spectrum for Proxima Centauri from the MUSCLES spectral survey \citep[][]{france_muscles_2016, youngblood_muscles_2016, loyd_muscles_2016} and extended to 10~$\mu$m using the spectrum from \citet{ribas_full_2017}. Radiative transfer through the atmosphere is treated by the Suite of Community Radiative Transfer codes based on Edwards and Slingo (SOCRATES) scheme \citep[][]{edwards_studies_1996}. The UM is one of the leading models in predicting the Earth's weather and climate and has been adapted for the study of several types of exoplanets, including terrestrial planets \citep[e.g.][]{mayne_using_2014, boutle_exploring_2017, lewis_influence_2018, yates_ozone_2020, eager_implications_2020, braam_lightning-induced_2022, ridgway_3d_2023} but also Mini-Neptunes \citep[e.g.][]{drummond_effect_2018} and hot Jupiters \citep[e.g.][]{mayne_unified_2014, mayne_results_2017}. Furthermore, the UM was part of the TRAPPIST-1e Habitable Atmosphere Intercomparison (THAI) project \citep[][]{turbet_trappist-1_2022, sergeev_trappist-1_2022, fauchez_trappist-1_2022}.

We use UKCA to simulate the 3-D atmospheric chemical composition, by including its description of gas-phase chemistry. UKCA is fully coupled to the UM for large-scale advection, convective transport and boundary layer mixing of the chemical tracers \citep[][]{morgenstern_evaluation_2009, oconnor_evaluation_2014, archibald_description_2020}. The Fast-JX photolysis scheme is implemented within UKCA, to calculate photolysis rates of chemical species in the atmosphere \citep[][]{wild_fast-j_2000, bian_fast-j2_2002, neu_global_2007, telford_implementation_2013}. By taking into account the varying optical depths of Rayleigh scattering, absorbing gases, and clouds from the UM, Fast-JX provides an interactive treatment of photolysis in calculating the 3-D distribution of chemical species in the atmosphere. We distribute the stellar flux from Proxima Centauri over the 18 wavelength bins of Fast-JX, as shown in \citet{braam_lightning-induced_2022} and their Figure 1. These fluxes are synchronised to the orbital distance of Proxima Centauri b which provides an interactive calculation of photolysis rates over the planetary orbit. The chemistry included is a reduced version of UKCA's Stratospheric-Tropospheric scheme \citep[StratTrop,][]{archibald_description_2020}, including the Chapman mechanism of ozone formation, and the hydrogen oxide (HO$_{\rm{x}}$=H+OH+HO$_2$) and nitrogen oxide (NO$_{\rm x}$=NO+NO$_2$) catalytic cycles. This results in 21 chemical species that are connected by 71 reactions. A full list of species and reactions can be found in the appendix of \citet{braam_lightning-induced_2022}.

\subsection{Metrics}\label{sec:metrics}
The meridional circulation is diagnosed using the mean meridional mass streamfunction (in kg~s$^{-1}$), which calculates the northward mass flux above pressure $P$:
\begin{equation}\label{eq:mm_stream}
    \Psi_m = \frac{2\pi R_p \cos{\phi}}{g}\int^P_0 \Bar{v}dP,
\end{equation}
with $R_p$ as the planetary radius, $g$ as the gravitational acceleration and $\Bar{v}$ as the zonal and temporal mean of the northward velocity component at latitude $\phi$. Earlier studies using this metric for synchronously rotating exoplanets \citep[e.g.][]{merlis_atmospheric_2010, edson_atmospheric_2011, carone_connecting_2015, haqq-misra_geothermal_2015, carone_stratosphere_2018} showed 1) the existence of tropospheric Hadley and Ferrel cells transporting heat and mass from the equatorial to polar regions and 2) the impact of orbital configuration on the Brewer-Dobson circulation in the stratosphere \citep[][]{carone_stratosphere_2018}. 

However, with the fixed substellar point of synchronously rotating planets, the mean meridional circulation varies depending on the position relative to the substellar point: for example, the hemispheric mean meridional circulation can vary significantly between the dayside and nightside. The zonal circulation is analogous to the Walker circulation cells on Earth, with rising motion at the location of the heat source, followed by eastward and westward flow aloft and, after descending on the nightside, a return flow along the surface back to the heat source \citep[][]{geisler_linear_1981}. The mean zonal mass streamfunction can be used to calculate the eastward mass flux above pressure $P$:
\begin{equation}\label{eq:zm_stream}
    \Psi_z = \frac{2\pi R_p}{g}\int^P_0 \Bar{u}dP,
\end{equation}
where $\Bar{u}$ is the meridional mean of the zonal velocity component. For slow rotators, the mean zonal circulation connects the substellar and antistellar points \citep[][]{gill_simple_1980, merlis_atmospheric_2010, showman_equatorial_2011, heng_atmospheric_2011, edson_atmospheric_2011, haqq-misra_geothermal_2015, haqq-misra_demarcating_2018}. The substellar-antistellar circulation also consists of a cross-polar flow \citep[][]{haqq-misra_geothermal_2015}.

\begin{figure}
\includegraphics[width=1\columnwidth]{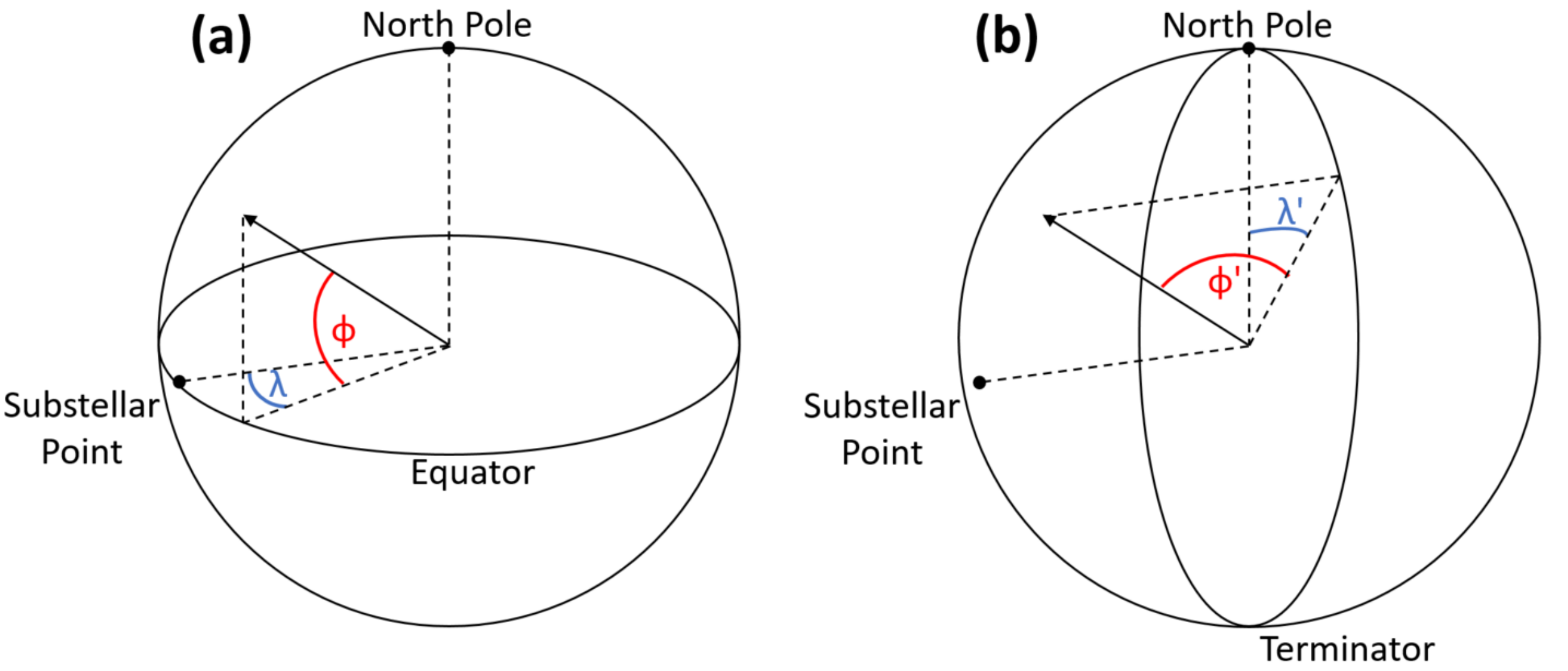}
\caption{Geographic coordinate system (a) showing latitude $\phi$ and longitude $\lambda$, with the substellar point located at (0$^\circ$,0$^\circ$). In the tidally-locked coordinate system (b) we use tidally-locked latitude $\phi'$ and tidally-locked longitude $\lambda'$ and the substellar point is located at $\phi'{=}$90$^\circ$. The nightside corresponds to negative $\phi'$. Figure adapted from \citet{koll_deciphering_2015}.}
\label{fig:tl_coords}
\end{figure}

As elaborated in Section~\ref{sec:intro}, the total wind flow on synchronously rotating exoplanets consists of several components. We perform a Helmholtz decomposition of the total wind flow, following \citet{hammond_rotational_2021}. This decomposes the total wind flow into its rotational, eddy rotational, and divergent components. The divergent wind mainly drives the substellar-antistellar overturning circulation \citep[][]{hammond_rotational_2021, sergeev_trappist-1_2022}. Since the divergent component is roughly isotropic around the substellar point, we can move from the usual latitude-longitude or geographic coordinate system to a tidally-locked coordinate system \citep[][]{koll_deciphering_2015, hammond_rotational_2021}. The transformation between geographic coordinates and tidally-locked coordinates is illustrated in Figure~\ref{fig:tl_coords}. The tidally-locked latitude $\phi'$ is measured as the angle from the terminator and the tidally-locked longitude $\lambda'$ is the angle about the substellar point, with the geographic North Pole located at ($\phi',\lambda'){=}(0,0)$ in tidally-locked coordinates. The substellar point and antistellar point correspond to $\phi'{=}90^\circ$ and ${-}90^\circ$, respectively. It was shown by \citet{hammond_rotational_2021} that integrating the continuity equation in tidally-locked coordinates over $\lambda'$ leads to the tidally-locked mean meridional mass streamfunction:
\begin{equation}\label{eq:mm_stream_tl}
    \Psi'_m = \frac{2\pi R_p \cos{\phi'}}{g}\int^P_0 \Bar{v'}dP,
\end{equation}
where $\Bar{v'}$ is the zonal mean of the meridional velocity component at tidally-locked latitude $\phi'$. In this system, the meridional mass streamfunction calculates the mass flux toward the antistellar point (along lines of constant $\lambda'$), connecting the substellar and antistellar points and also taking cross-polar flow into account.

Since we are particularly interested in the transport of ozone around the planet, we weight the stream functions using the ozone mass mixing ratio ($\chi_{\rm{O3}}$), which is measured as the mass of ozone per unit mass of air in a parcel. This gives us the ozone mass streamfunction:
\begin{equation}\label{eq:mm_stream_tl_o3}
    \Psi'_{\rm{O_3}} = \Psi'\times\chi_{\rm{O_3}},
\end{equation}
which can be applied generally using any of the streamfunctions in Equations~\ref{eq:mm_stream}, \ref{eq:zm_stream} or \ref{eq:mm_stream_tl} to give the ozone-weighted meridional, zonal, or the tidally-locked meridional mass streamfunction.

\subsection{Experimental Setup}\label{sec:setup}
We use the final state of the `Chapman+HO$_x$+NO$_x$' simulation from \citet{braam_lightning-induced_2022} for the analysis. The atmosphere was initialized at an Earth-like atmospheric composition, using preindustrial values of N$_2$, O$_2$ and CO$_2$ \citep[see also][]{boutle_exploring_2017}. Water vapour profiles are interactively determined by evaporation from the slab ocean. The HO$_{x}$ and NO$_{x}$ species are initialized at mass mixing ratios of 10$^{-9}$ and 10$^{-15}$, respectively. We report results from our simulation as 600-day mean of the CCM output (equal to ${\sim}$50 orbits of Proxima Centauri b) after spinning up for 20 Earth years, to ensure the simulation has reached a dynamical and chemical steady state. The dynamical steady state was determined by the stabilisation of the surface temperature and radiative balance at the top of the atmosphere. The chemical steady state was determined by the stabilisation of ozone as a long-lived species, through the total column and volume mixing ratios. In diagnosing the impact of dynamical processes on the ozone distribution, parts of the spin-up period have also been used to plot the evolution of chemically inert tracers (see Figure~\ref{fig:ageair_tllat} below). The analysis of temporal variability in Section~\ref{sec:tempevol} is based on a 6-day output over 900 days of simulation after reaching a steady state, to ensure we include potential variability at longer timescales.

\section{Results}\label{sec:results}
In this section, we start with a brief description of the planetary climate and ozone layer. After that, we discuss the atmospheric circulation followed by its impact on the distribution of ozone around the planet, elaborating on the stratospheric overturning circulation. Lastly, we perform a comparison of relevant lifetimes in the atmosphere.

\subsection{Planetary climate and atmospheric ozone}\label{subsec:clim_o3}
\begin{figure}
\includegraphics[width=\columnwidth]{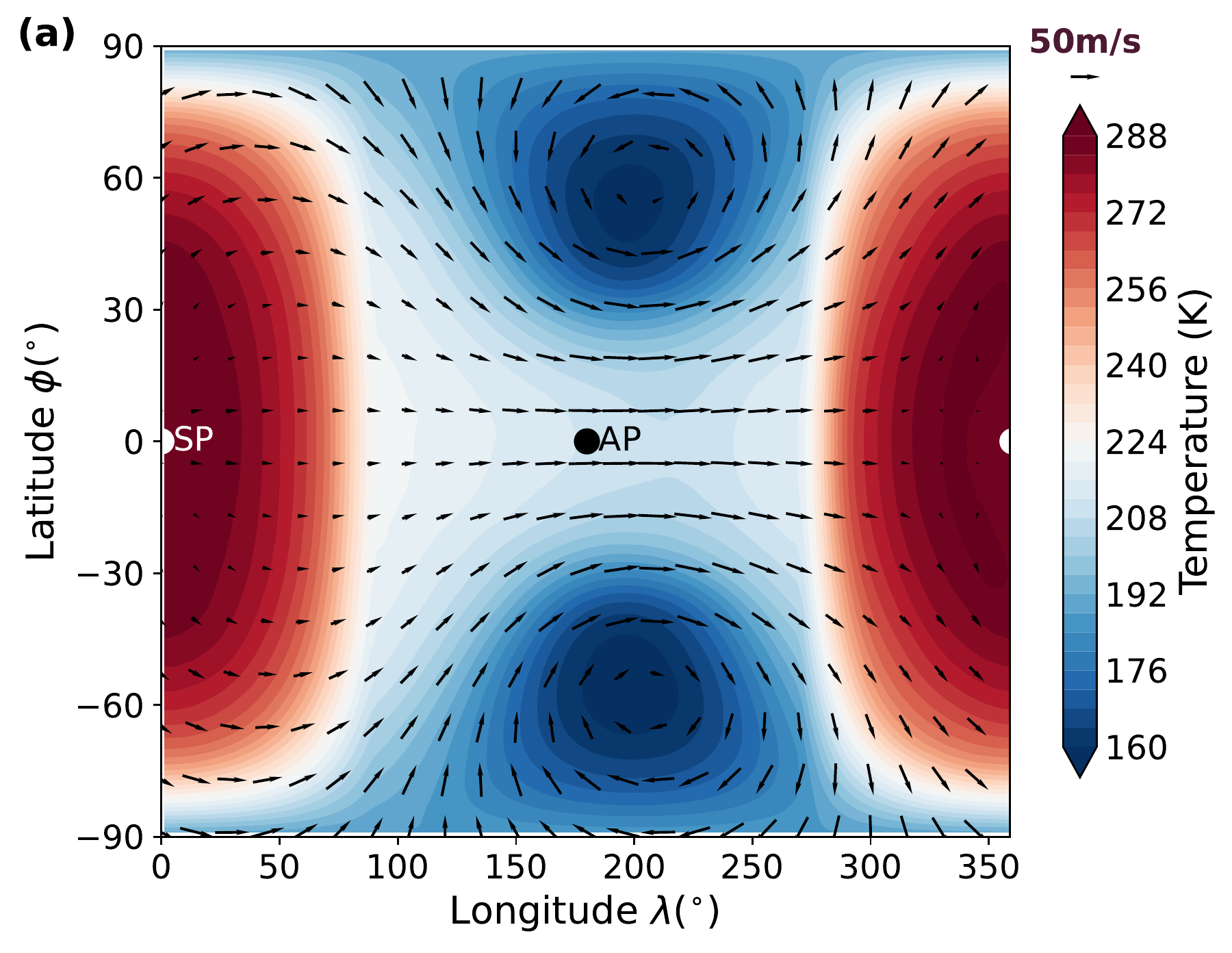}
\includegraphics[width=\columnwidth]{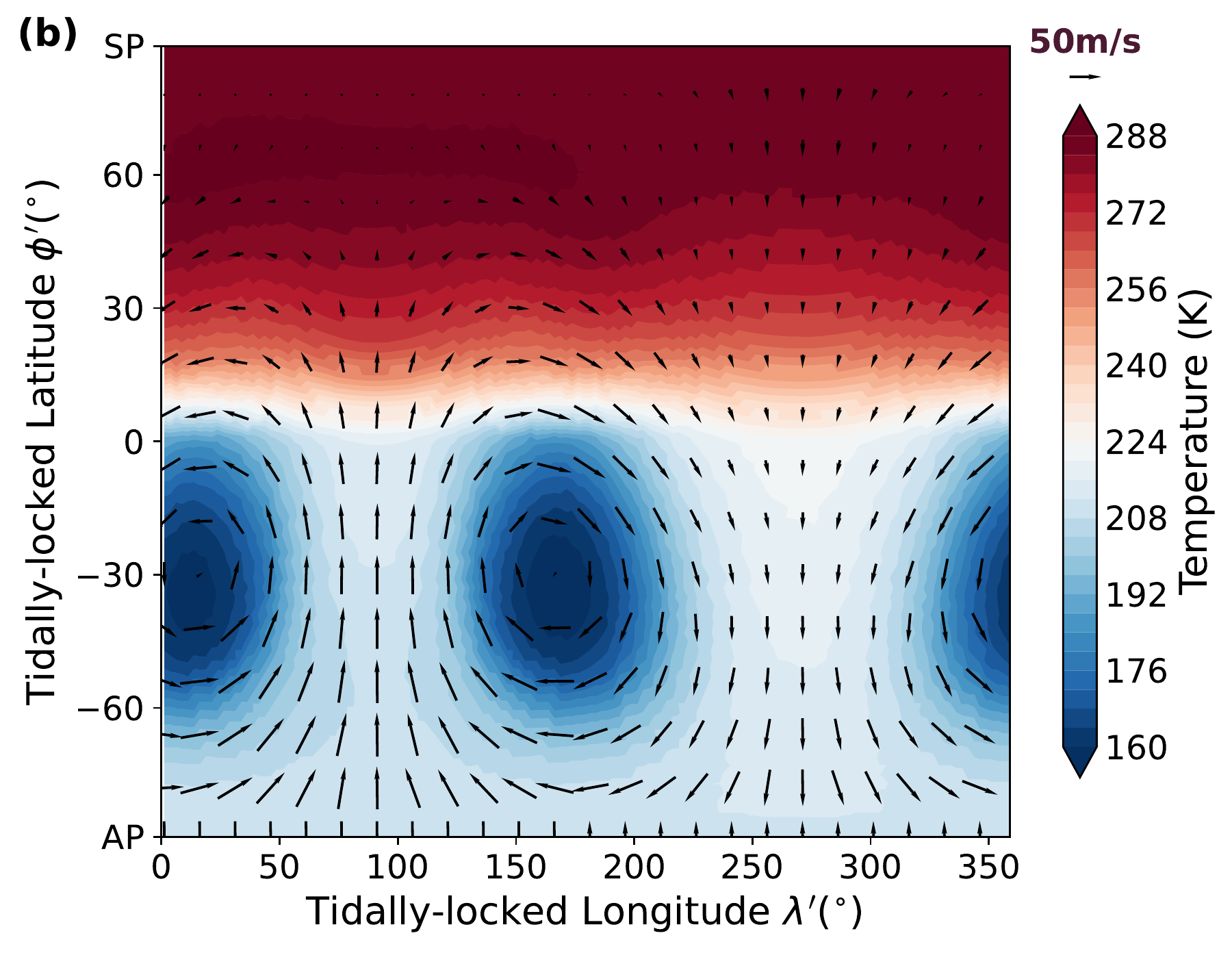}
\caption{Temporal mean surface temperature over 50 orbits of Proxima Centauri b, using (a) the geographic coordinate system and (b) the tidally-locked coordinate system \citep{koll_deciphering_2015}. The substellar point is transformed from ($\phi,\lambda){=}(0^\circ,0^\circ)$ in geographic coordinates (white dot) to $\phi'{=}90^\circ$ in tidally-locked coordinates, as also shown in Figure~\ref{fig:tl_coords}. Overplotted are the horizontal wind vectors at P${\approx}400$~hPa, showing both the tropospheric jet and the existence of the Rossby gyres on the nightside.}
\label{fig:pcb_temp}
\end{figure}

The simulated climate of Proxima Centauri b is broadly similar to that described by \citet{boutle_exploring_2017}. Furthermore, the formation of an ozone layer under quiescent stellar radiation is explained in detail by \citet{yates_ozone_2020} and \citet{braam_lightning-induced_2022}. Here, we give a brief description of the details essential for this study. The simulated surface temperature of Proxima Centauri b is shown in Figure~\ref{fig:pcb_temp}, using a geographic coordinate system in panel (a) and tidally-locked coordinate system in panel (b). Both panels show the dayside-to-nightside contrast characteristic of synchronous rotation, with dayside maxima in surface temperature of up to 289~K and minima of 157~K over the nightside Rossby gyres. Figure~\ref{fig:pcb_temp}b demonstrates the usefulness of the tidally-locked coordinate system in identifying the dayside-to-nightside contrasts, with the terminator located at $\phi'{=}0^\circ$. The horizontal wind vectors are shown at P${\approx}400$~hPa, illustrating the tropospheric jet as well as the Rossby gyres on the nightside. The dayside-to-nightside circulation is part of an overturning circulation across multiple pressure levels that will be described in more detail in Section~\ref{subsec:circulation}. At the locations of the nightside Rossby gyres \citep[][]{showman_matsuno-gill_2010}, we see the coldest areas on the planetary surface with air that is trapped and subject to radiative cooling. The atmospheric pressure in the gyres is relatively low, like the eye of tropical cyclones \citep[][]{schubert_distribution_2007}. The gyres are relatively isolated from the rest of the hemisphere and their edges act as mixing barriers \citep{vallis_atmospheric_2017}. The gyres are a general feature of slowly rotating exoplanets in a synchronous orbit that have a single equatorial jet in the troposphere \citep[][]{sergeev_bistability_2022}.

\begin{figure}
\includegraphics[width=\columnwidth]{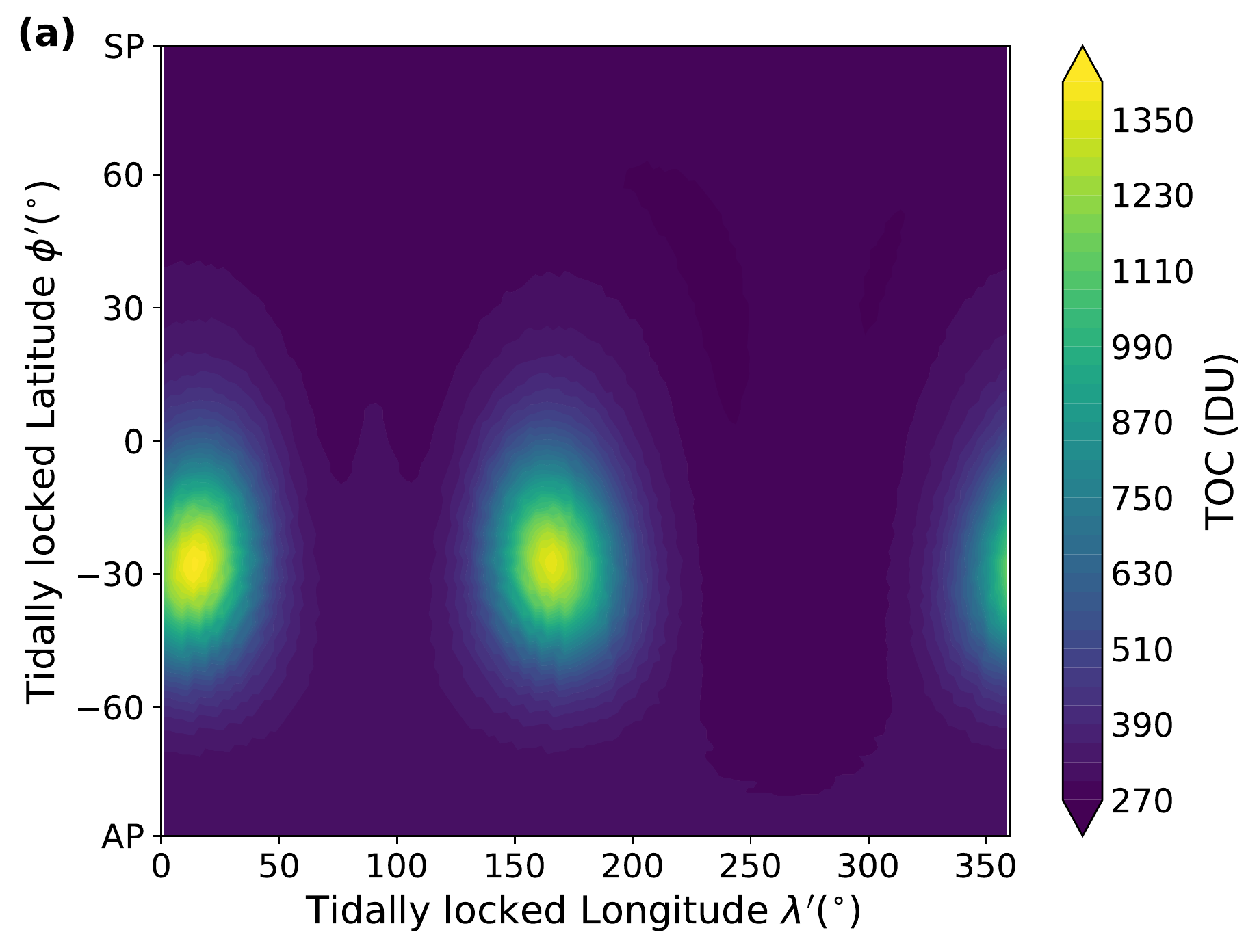}
\includegraphics[width=\columnwidth]{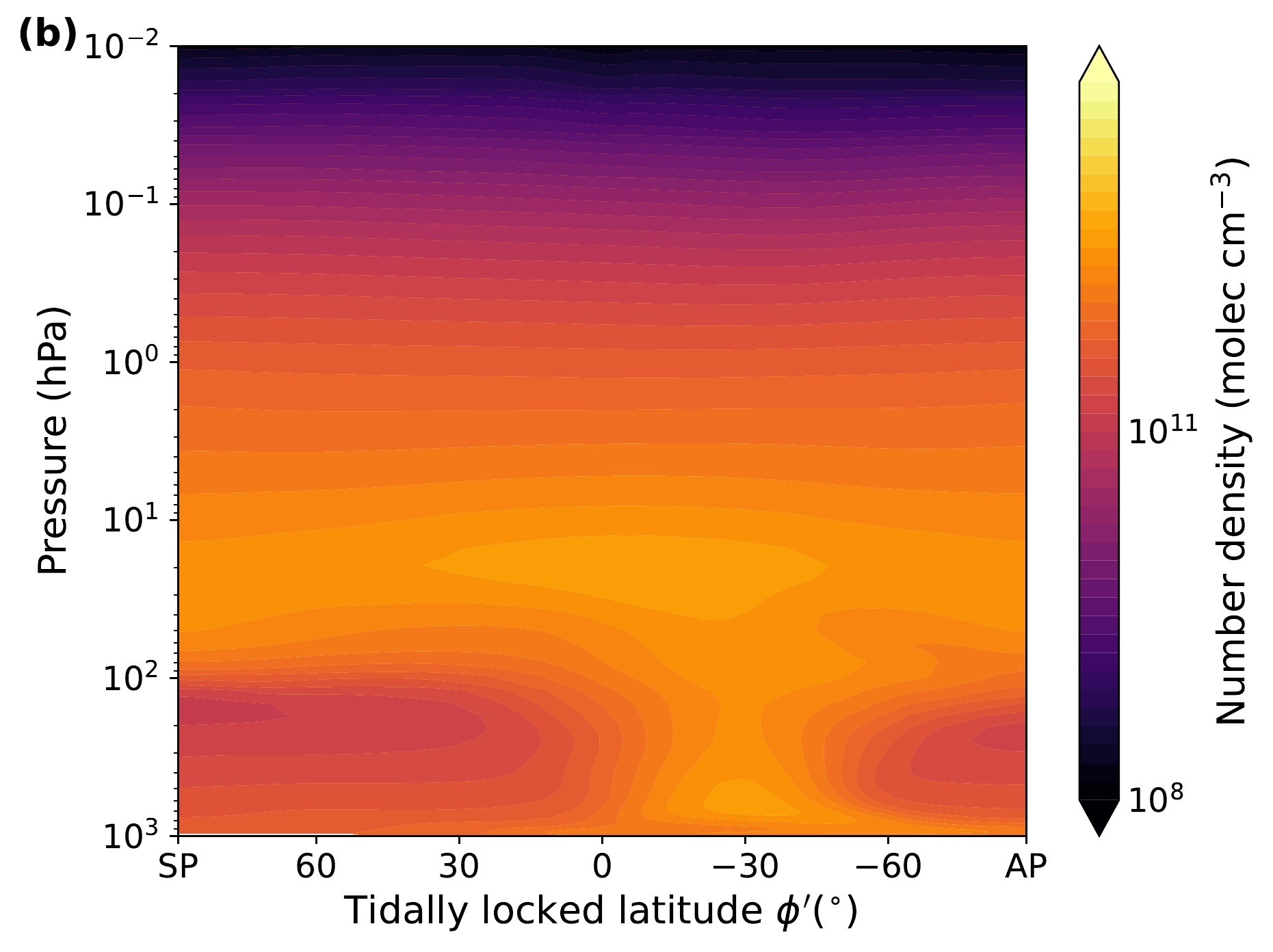}
\caption{(a) Total ozone column and (b) meridional mean ozone number density, both taking means over 50 orbits of Proxima Centauri b. Both plots illustrating the spatially variable ozone layer with accumulation at the locations of the nightside Rossby gyres ($-60{<}\phi'{<}0$).}
\label{fig:pcb_o3distrib}
\end{figure}

We find a spatially variable distribution of ozone in Figure~\ref{fig:pcb_o3distrib}a, with a relatively thin dayside ozone layer and accumulation of ozone on the nightside. Typical values for the vertically-integrated ozone column on Earth's are 200--400 Dobson Units (DU: 1~DU${=}2.687\times10^{20}$~molecules~m$^{-2}$), with lower values over the equatorial regions and ozone hole and higher values over high-latitude regions \citep[][]{eyring_long-term_2013}. For synchronously rotating planets, most of the dayside ozone column falls within this range. The locations of the nightside Rossby gyres correspond to the maxima in the thickness of the ozone column, reaching up to 1401~DU. The gyres are not fully symmetric, evident from slightly different shapes and the average ozone columns: the area-weighted mean column of the low-$\lambda'$ gyre (for $\lambda'{\leq}70$ and $\lambda'{>}320^\circ$) is equal to 626~DU and of the mid-$\lambda'$ gyre ($110{<}\lambda'{\leq}220^\circ$) to 601~DU, both confined between tidally-locked latitudes ${-}60{<}\phi'{<}0^\circ$. Figure~\ref{fig:pcb_o3distrib}b shows that the accumulation of ozone at the gyre locations mostly occurs in the lower atmosphere, at pressure levels corresponding to the troposphere (${<}100$~hPa).

\begin{figure}
\includegraphics[width=\columnwidth]{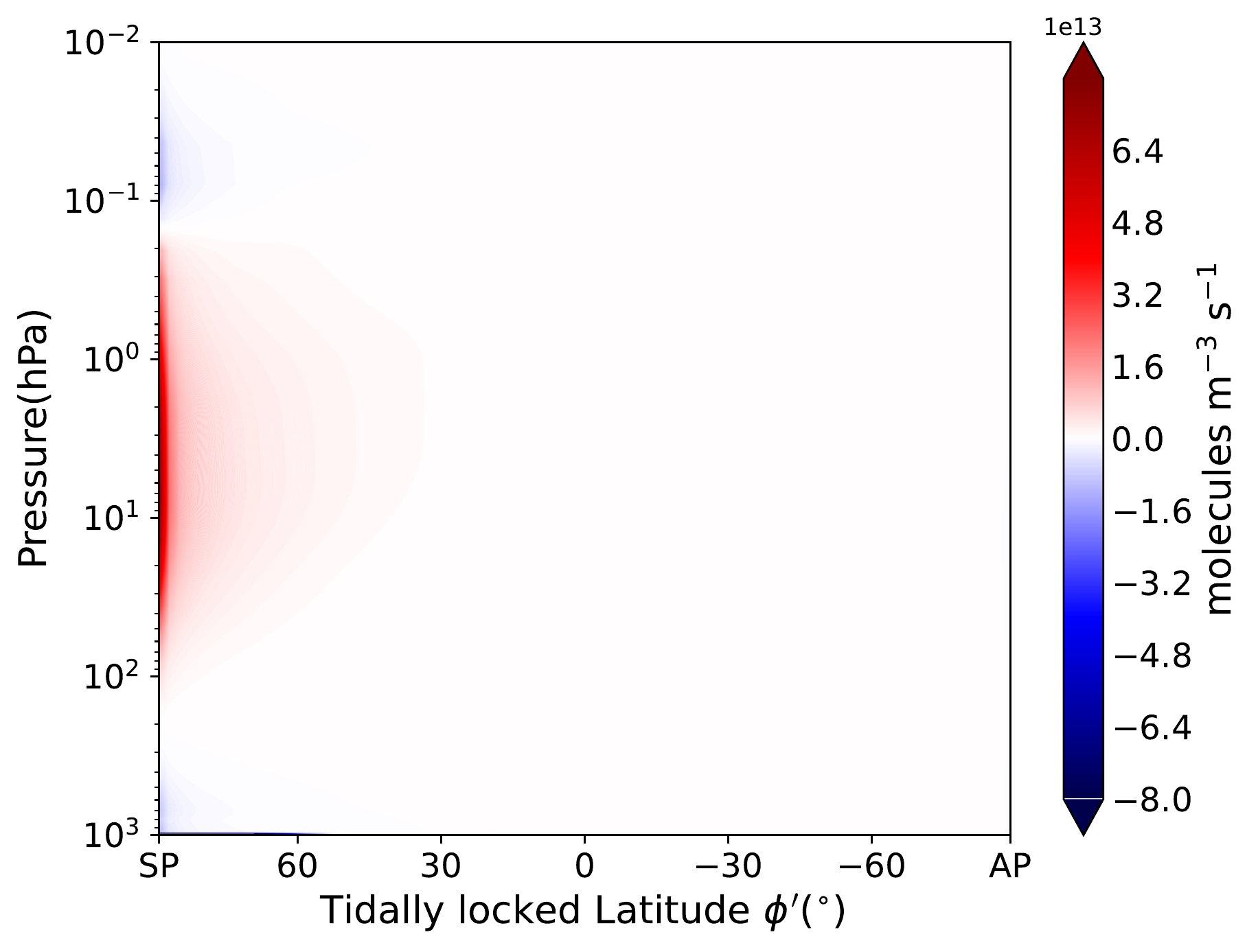}
\caption{Meridional mean ozone chemical tendency (production-loss) in tidally-locked coordinates, showing that ozone production is limited to the planet's dayside.}
\label{fig:pcb_o3tendency}
\end{figure}

The existence of such a spatially variable ozone layer depends on a complex interplay between photochemistry and atmospheric dynamics and changes as a function of incoming stellar radiation and planetary rotation state \citep[][]{chen_habitability_2019, chen_persistence_2021}. The production mechanisms for atmospheric ozone are relatively well-understood and due to photochemistry: in the presence of stellar radiation molecular oxygen will dissociate and form ozone through the Chapman mechanism \citep[][]{chapman_xxxv_1930}. The 3-D impact of M-dwarf radiation on the Chapman mechanism has been explored by previous studies, both in quiescent \citep[][]{yates_ozone_2020, braam_lightning-induced_2022} and flaring conditions \citep[][]{ridgway_3d_2023}. In all cases, an ozone layer develops around the planet. As such exoplanets are likely to rotate synchronously around their host star \citep[][]{barnes_tidal_2017}, stellar radiation and the photochemical production of ozone are limited to the planetary dayside. This is illustrated in Figure~\ref{fig:pcb_o3tendency}, showing the time-averaged chemical tendency of ozone. The tendency denotes the balance between the production and loss of ozone due to chemical processes. We find that ozone production mainly occurs at high $\phi'{>}40^\circ$ (i.e., close to the substellar point), whereas ozone production is practically absent at the locations of the nightside gyres (${-}60{<}\phi'{<}0^\circ$). Hence, another mechanism must be driving the relatively enhanced ozone abundances at the locations of the nightside Rossby gyres.

\subsection{Overturning circulations}\label{subsec:circulation}
\begin{figure*}
\includegraphics[width=0.66\columnwidth]{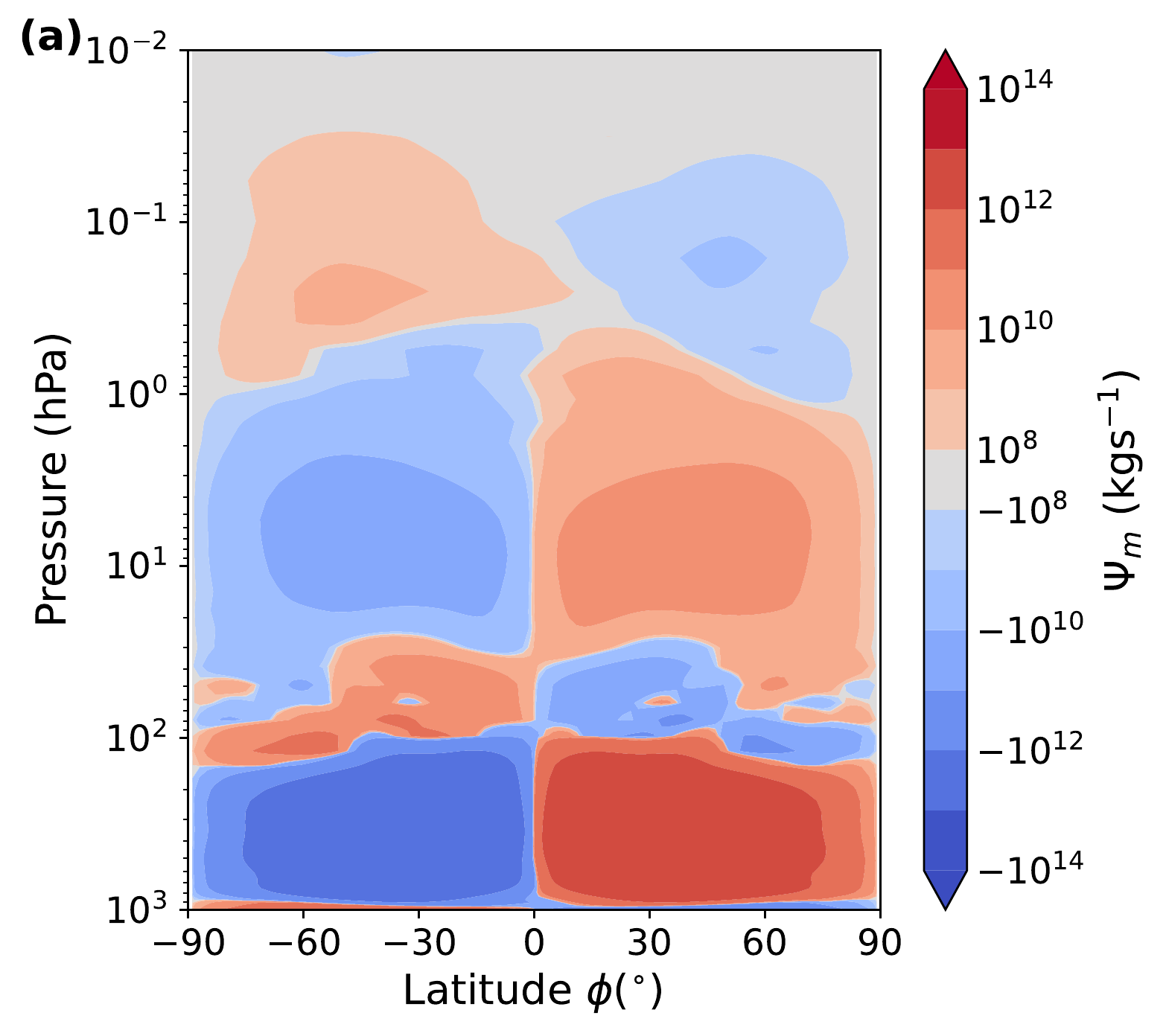}
\includegraphics[width=0.66\columnwidth]{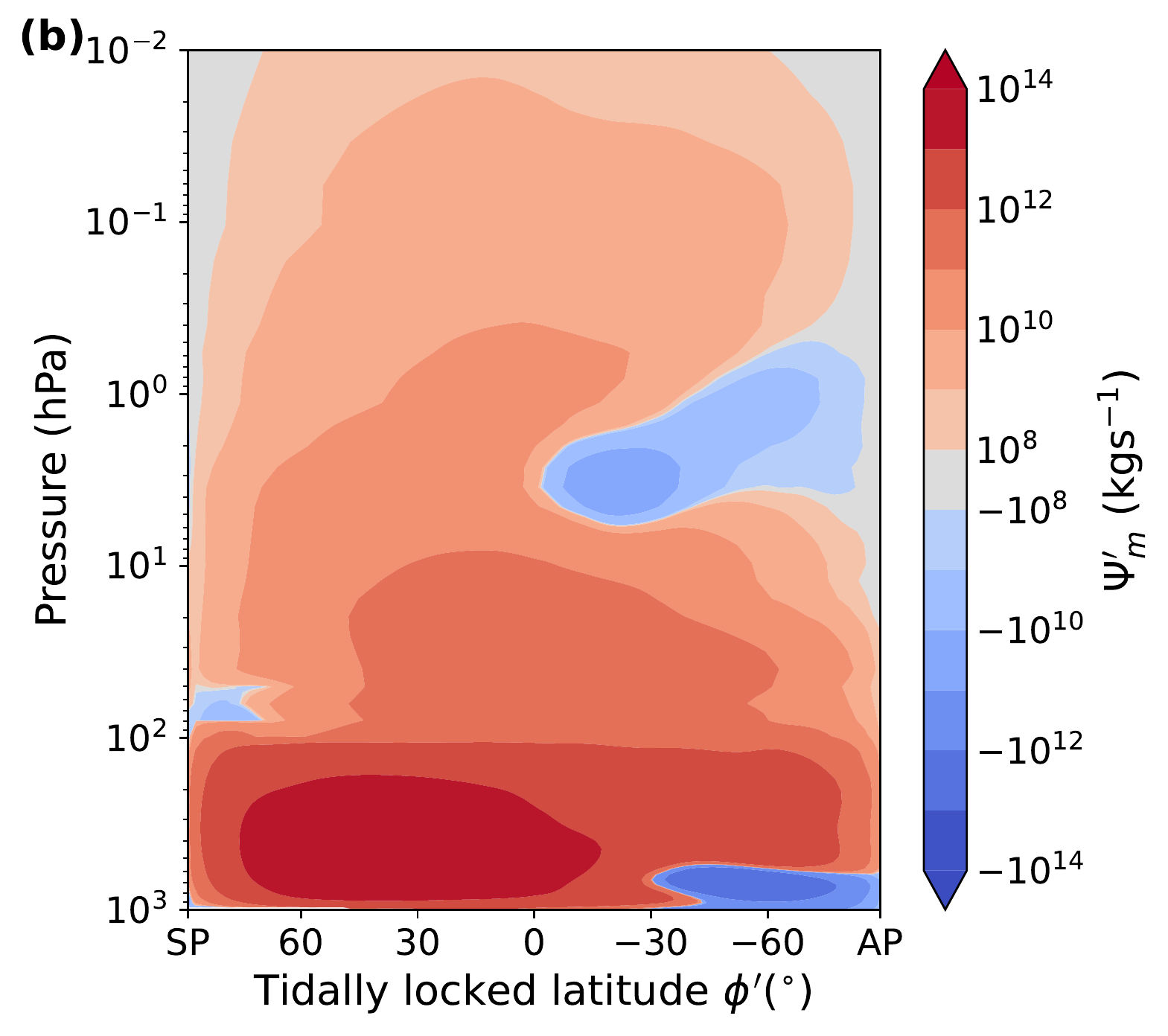}
\includegraphics[width=0.66\columnwidth]{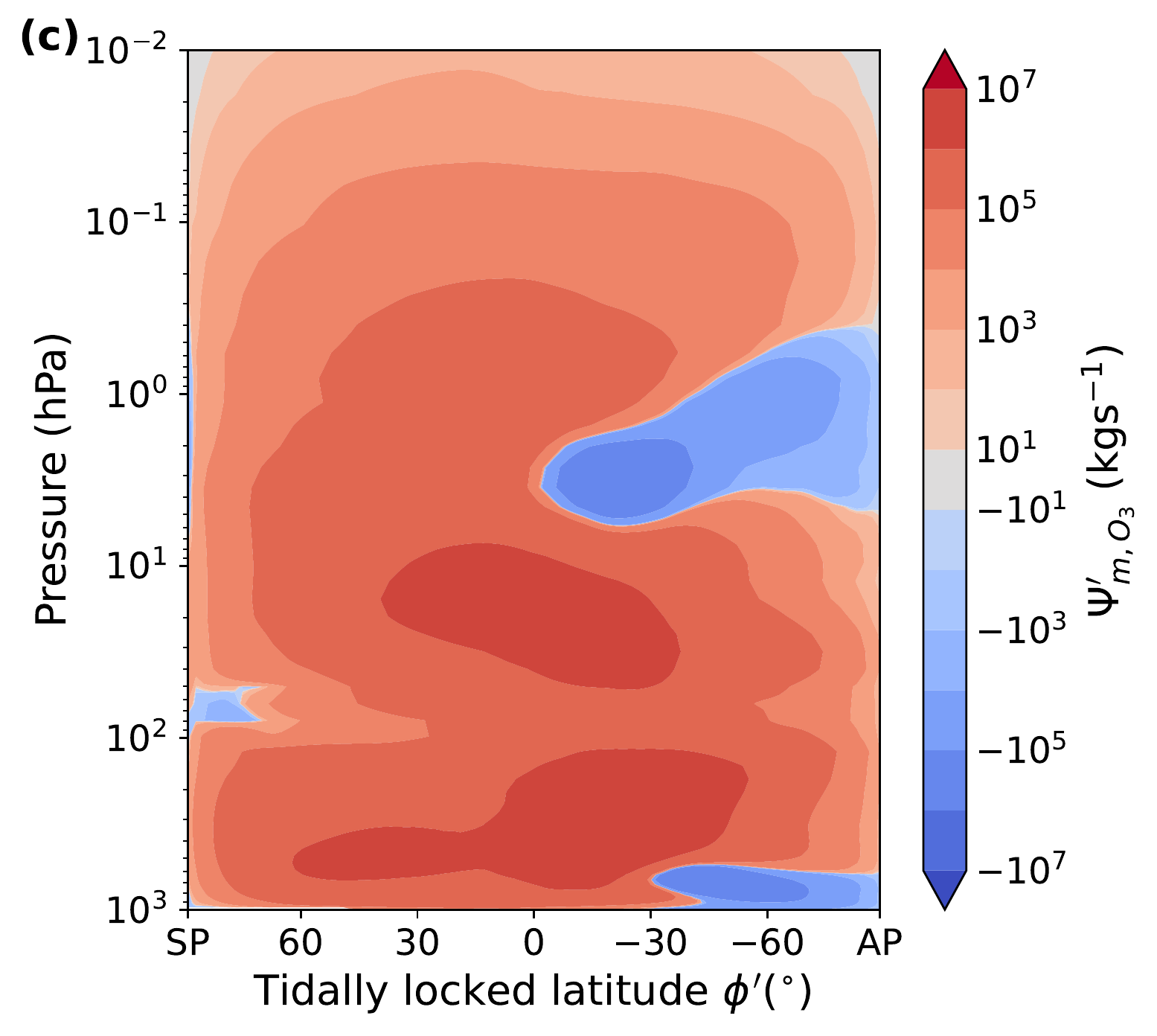}
\caption{Zonal mean meridional mass streamfunctions illustrating different aspects of atmospheric circulation. Positive values (red) indicate clockwise and negative values (blue) anticlockwise motion. (a) The meridional mass streamfunction in geographic coordinates $\Psi_m$ (Equation~\ref{eq:mm_stream}) shows equator-to-pole stratospheric transport like the Brewer-Dobson circulation. (b) The meridional mass streamfunction in tidally-locked coordinates $\Psi'_m$ (Equation~\ref{eq:mm_stream_tl}) shows an overturning dayside-to-nightside circulation including a strong stratospheric component (above ${\sim}100$~hPa). (c) The meridional ozone mass streamfunction $\Psi'_{m,\rm{O_3}}$ (Equation~\ref{eq:mm_stream_tl_o3}) shows that this stratospheric component is significant in terms of transporting ozone from the dayside to the nightside.}
\label{fig:streamfunctions}
\end{figure*}

The relationship between the ozone distribution in Figure~\ref{fig:pcb_o3distrib} and the global atmospheric circulation becomes clear through the mass streamfunctions, as defined in Section~\ref{sec:metrics}. From left to right, Figure~\ref{fig:streamfunctions} shows the mean meridional mass streamfunctions $\Psi_m$, $\Psi'_m$ and $\Psi'_{m,\rm{O_3}}$ that have been calculated from the divergent wind component. A positive streamfunction (red contours) indicates clockwise circulation, and a negative streamfunction (blue) indicates anticlockwise circulation.

From Figure~\ref{fig:streamfunctions}a, we find strong poleward transport of air at tropospheric pressures (${>}100$~hPa) in a single thermally driven circulation cell \citep[][]{merlis_atmospheric_2010, edson_atmospheric_2011}. Moving up into the stratosphere, we find stacked layers of clockwise and anticlockwise circulation. The existence of poleward transport between ${\sim}50$ and ${\sim}1.5$~hPa indicates additional thermally-driven circulation cells. These cells transport aerosols and chemical tracers such as ozone from the equator to the poles through the stratosphere \citep[][]{carone_stratosphere_2018, chen_habitability_2019}. This equator-to-pole transport leads to an enhanced high latitude ozone layer on the dayside in geographic coordinates, with mean ozone columns of ${\sim}$490~DU above 80$^\circ$ North and South as compared to a mean of ${\sim}$290~DU between 10$^\circ$ North and 10$^\circ$ South \citep[see also][]{yates_ozone_2020, braam_lightning-induced_2022}. Since the stellar radiation at the poles is too weak to initiate the photochemistry responsible for ozone production, this polar enhancement has to be due to the poleward transport of ozone produced in the equatorial regions. 

Moving to tidally-locked coordinates using $\Psi'_m$ in Figure~\ref{fig:streamfunctions}b, we find a single overturning circulation cell that dominates the troposphere and transports air and heat from the dayside towards the nightside. A weaker anticlockwise circulating cell is present between the antistellar point and $\phi'{\approx}{-}30^\circ$, induced by the temperature gradient between those two points. The absence of anticlockwise motion when moving to lower pressure levels in Figure~\ref{fig:streamfunctions}b indicates that a connection between the tropospheric cell and the stratospheric circulation exists. An overturning circulation covers essentially all of the stratosphere, connecting the dayside and nightside. Air ascends in the ozone production regions (between $0.2$ and $100$~hPa, see Figure~\ref{fig:pcb_o3tendency}) and moves through the stratosphere towards the nightside, where it subsides at the locations of the nightside gyres and thus the locations of ozone accumulation as shown in Figure~\ref{fig:pcb_o3distrib}.

To quantify the impact of this mass transport on the distribution of ozone, we calculate the tidally-locked ozone-weighted mass streamfunction $\Psi'_{m,\rm{O_3}}$ (Equation~\ref{eq:mm_stream_tl_o3}) as shown in Figure~\ref{fig:streamfunctions}c. From the ozone mass streamfunction we infer that the circulation of ozone through the stratosphere provides a significant contribution to the dayside-to-nightside transport. The downward ozone transport at the $\phi'$ of the Rossby gyres (${-}60{<}\phi'{<}0^\circ$) indicates that this stratospheric dayside-to-nightside circulation drives ozone-rich air into the Rossby gyres and thus leads to ozone maxima on the nightside. 

\begin{figure*}
\includegraphics[width=0.8\columnwidth]{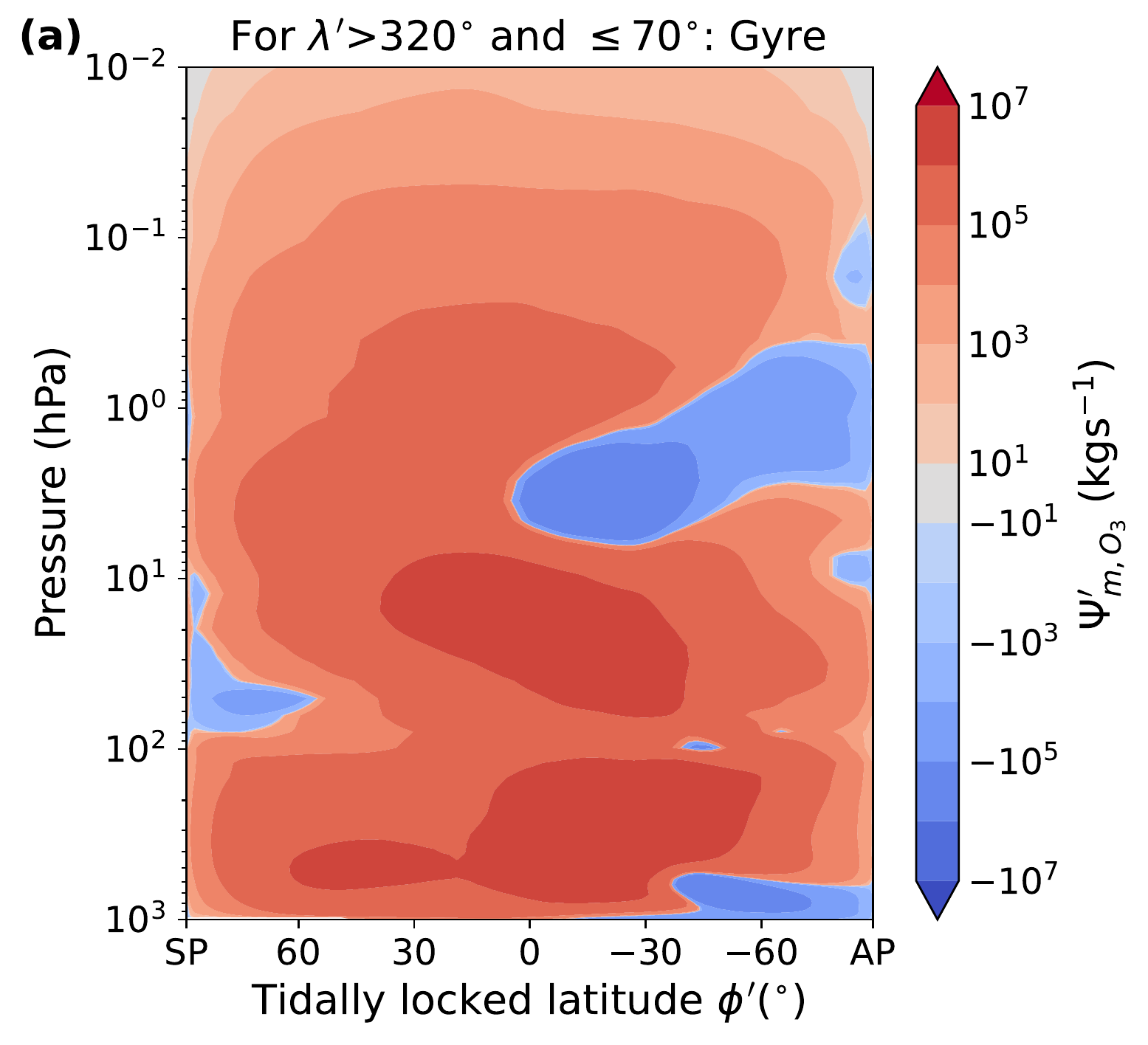}
\includegraphics[width=0.8\columnwidth]{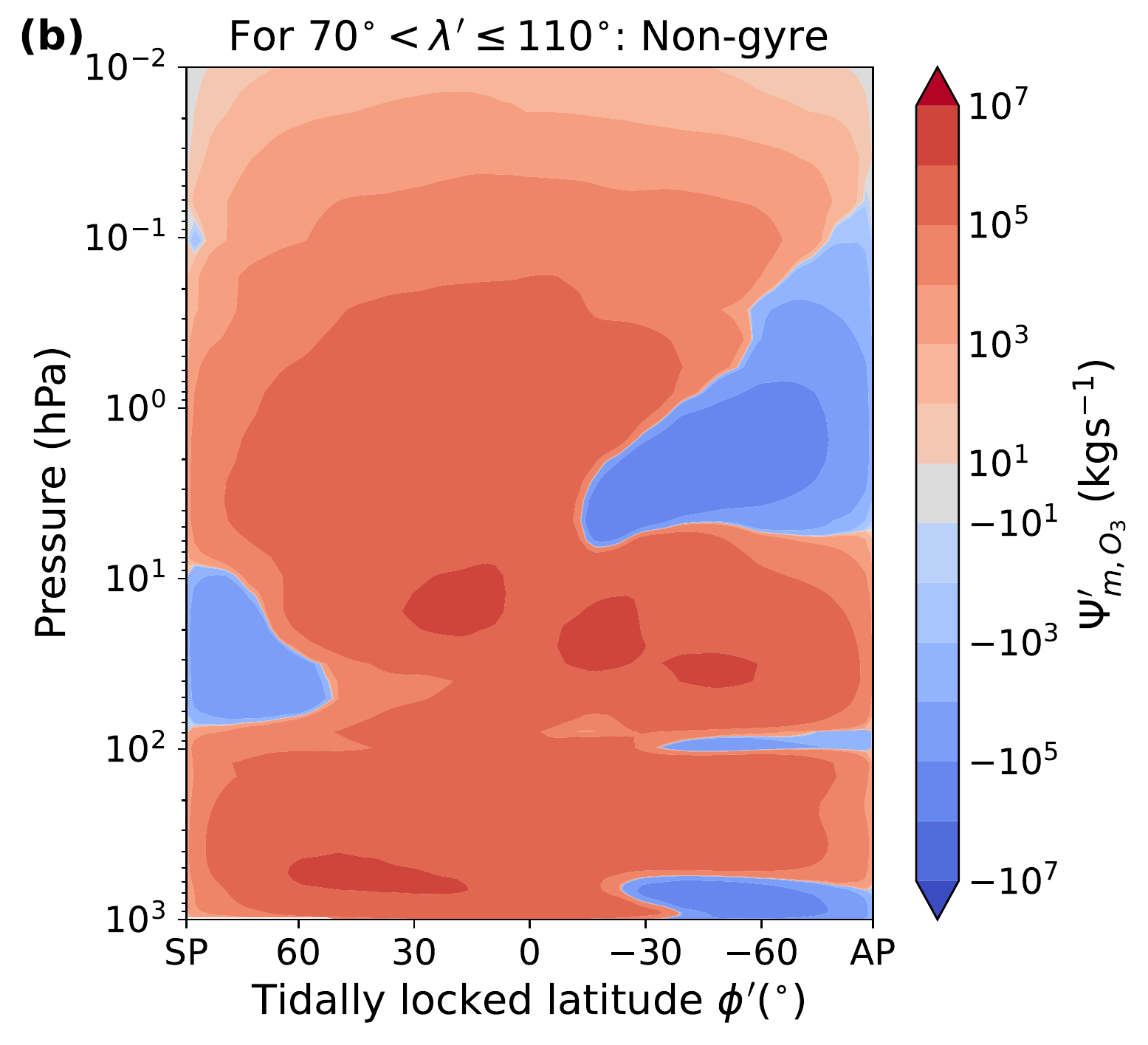}
\includegraphics[width=0.8\columnwidth]{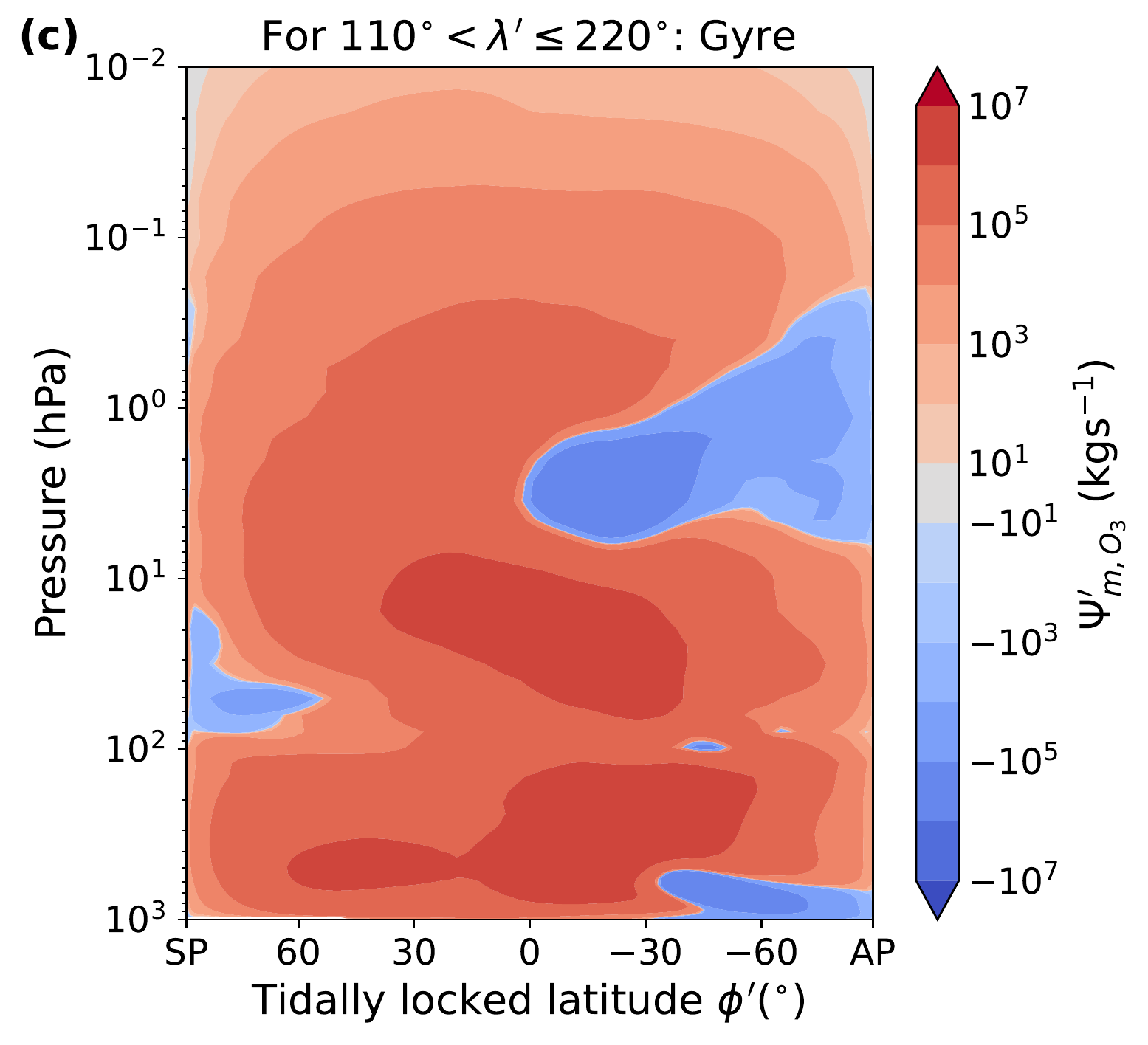}
\includegraphics[width=0.8\columnwidth]{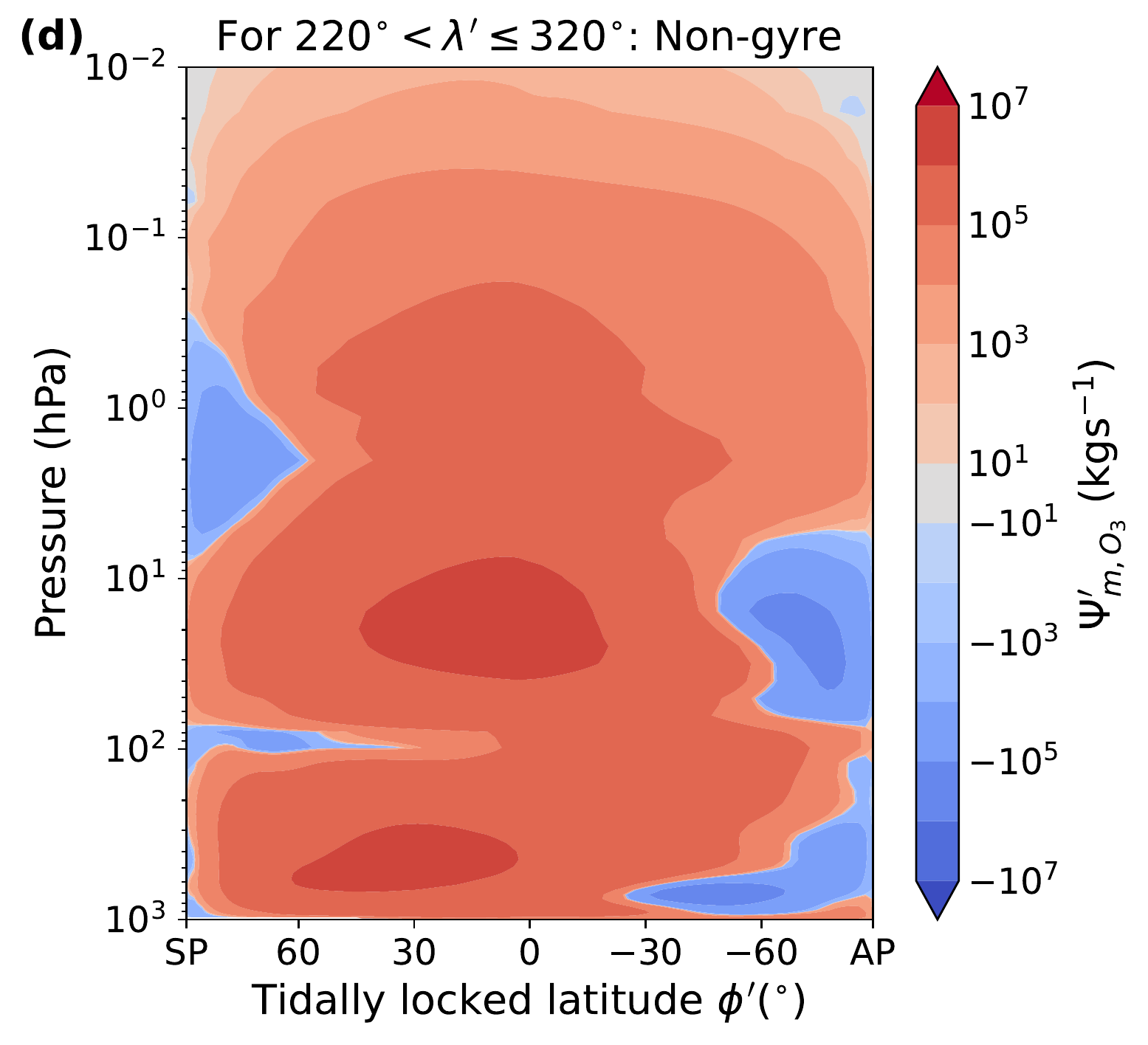}
\caption{The zonal mean meridional ozone mass streamfunction $\Psi'_{m,\rm{O_3}}$ (Equation~\ref{eq:mm_stream_tl_o3}) in tidally-locked coordinates, for ranges of tidally-locked longitude $\lambda'$ as shown by the titles of each of the four panels. Panels a and c denote $\lambda'$-ranges corresponding to the locations of the ozone accumulation in the Rossby gyres, following the distribution of ozone in Figure~\ref{fig:pcb_o3distrib}. The $\lambda'$-ranges in panels b and d correspond to the regions containing the superrotating jet. As such, panels a and c map out the meridional extent of the transport of ozone-rich air to the nightside.}
\label{fig:streamfunctions_long}
\end{figure*}

Figure~\ref{fig:streamfunctions_long} again shows $\Psi'_{m,\rm{O_3}}$, now separated into 4 ranges of $\lambda'$. Each of these $\lambda'$ ranges corresponds to a distinct feature of the ozone distribution in Figure~\ref{fig:pcb_o3distrib}a. Figure~\ref{fig:streamfunctions_long}a shows the $\lambda'$-range that contains the low-$\lambda'$ gyre ($\lambda'{>}320^\circ$ and $\lambda'{\leq}70^\circ$), and we can identify the dayside-to-nightside transport of ozone-rich air, followed by descending motion at $\phi'$ corresponding to the location of the Rossby gyres. The ozone is supplied from part of its production region (see Figure~\ref{fig:pcb_o3tendency}) between pressures of $0.3$~hPa and $20$~hPa. Figure~\ref{fig:streamfunctions_long}b shows the low-$\lambda'$-range that does not contain the gyres and instead includes the nightside-to-dayside component of the equatorial jet. $\Psi'_{m,\rm{O_3}}$ shows that there is a stratospheric clockwise circulation, but that this is separated from the lower parts of the atmosphere by an anticlockwise circulation at the $\phi'$ corresponding to the Rossby gyres and misses part of the ozone production regions between 10 and 100~hPa. Therefore, for $70{<}\lambda'{\leq}110^\circ$, no ozone accumulation is found following the stratospheric overturning circulation. Figure~\ref{fig:streamfunctions_long}c again indicates dayside-to-nightside transport of ozone-rich air, with ozone for the mid-$\lambda'$ gyre ($110{<}\lambda'{\leq}220^\circ$) being supplied from the ozone production regions between pressures of $0.3$~hPa and $15$~hPa. Lastly, Figure~\ref{fig:streamfunctions_long}d shows that in the final non-gyre range ($220{<}\lambda'{\leq}320^\circ$) there is a stratospheric overturning circulation transporting ozone-rich air, but this circulation misses part of the ozone production region between 0.3 and 10~hPa and is generally weaker than for the ranges containing the gyres. Furthermore, the air that descends below ${\sim}10$~hPa will meet the equatorial jet, leading to chemical destruction of ozone (due to HO$_\mathrm{x}$-rich air from the dayside) or advection back to the dayside followed by photochemical destruction. Therefore, this $\lambda'$-range is not accumulating ozone in the lower part of the atmosphere.

\begin{figure}
\includegraphics[width=1\columnwidth]{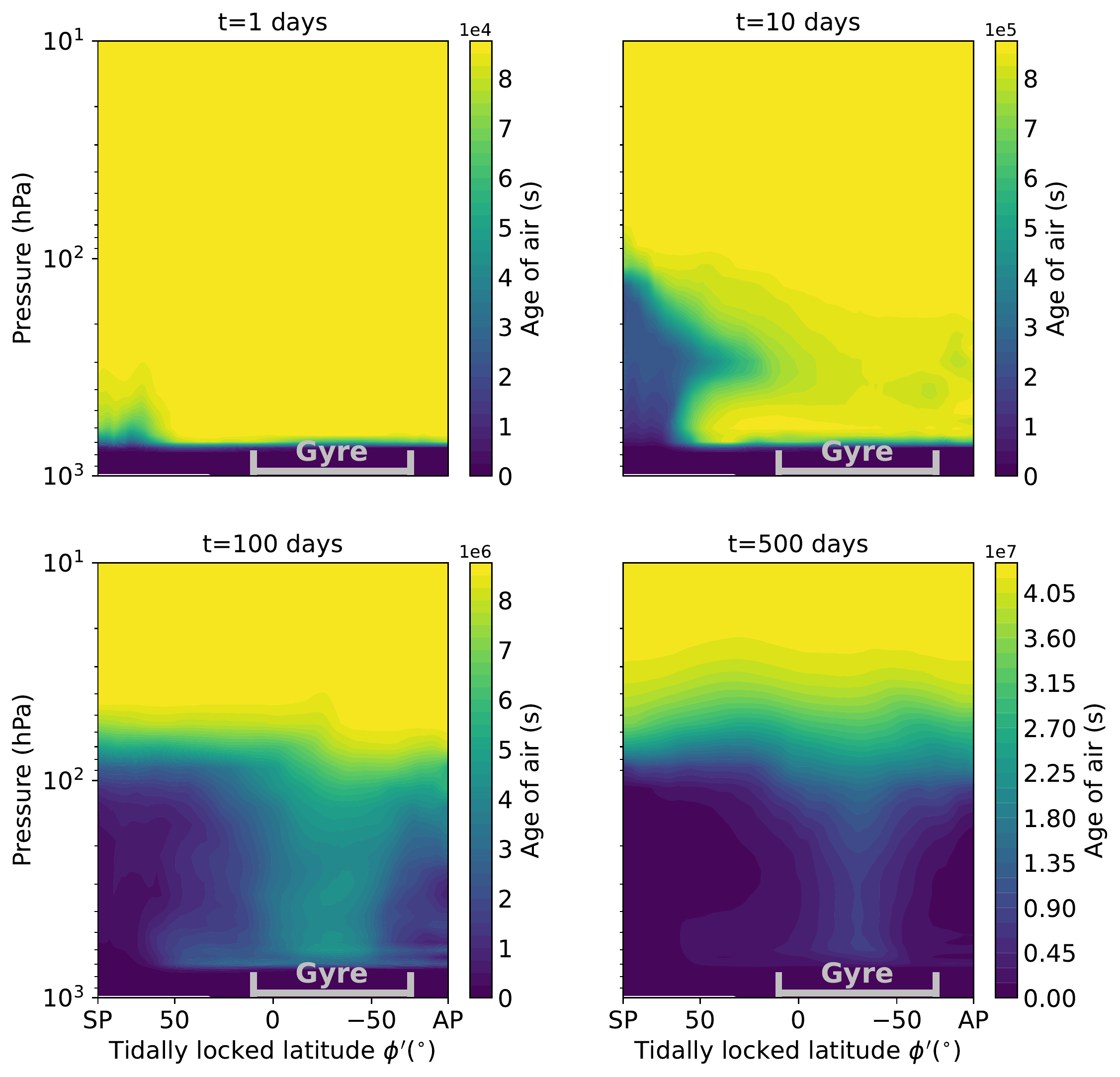}
\caption{Age-of-air tracer during the spin-up of the simulation, showing the mean meridional distribution in tidally-locked coordinates. As a passive tracer, it is only affected by dynamical processes (advective and convective). As such, the age-of-air measures the time it takes a parcel to rise from the lowest atmospheric layers (at ${\sim}2$~km or 700~hPa) into the stratosphere. The tracer values are reset to 0 in the lowest atmospheric layers at every model timestep. We also show the tidally-locked latitudes corresponding to the nightside gyres in grey.}
\label{fig:ageair_tllat}
\end{figure}

Our interpretation of the atmospheric dynamics is supported by an age-of-air tracer experiment. In Figure~\ref{fig:ageair_tllat}, we show the zonally-averaged time evolution of the age-of-air-tracer during the model spin-up period. As a passive tracer, it is only affected by dynamical processes in the UM, including both advection and convection. The age-of-air tracer is initialised at 0~s and provides a measure of the amount of time that has passed since an air parcel was last found in the lowest layers of the atmosphere (below ${\sim}2$~km or 700~hPa). As such, the tracer measures the time it takes a parcel to rise from these lowest layers into the stratosphere. The tracer values are reset to 0 in the lowest layers at every model timestep. With the evolution of the age-of-air tracer over $\phi'$ in Figure~\ref{fig:ageair_tllat} we show that air rises over and around the substellar point, already providing much younger air to the dayside troposphere (${<}15$~km) after 10 days of simulation. After 100 days, we find that most of the troposphere has been replenished with much younger air, except for the nightside gyres between ${-}60^\circ{<}\phi'{<}0^{\circ}$. This picture persists after 500 days, showing that the age-of-air tracer in the nightside gyres is fed by older air from the stratosphere.

\begin{figure}
\includegraphics[width=\columnwidth]{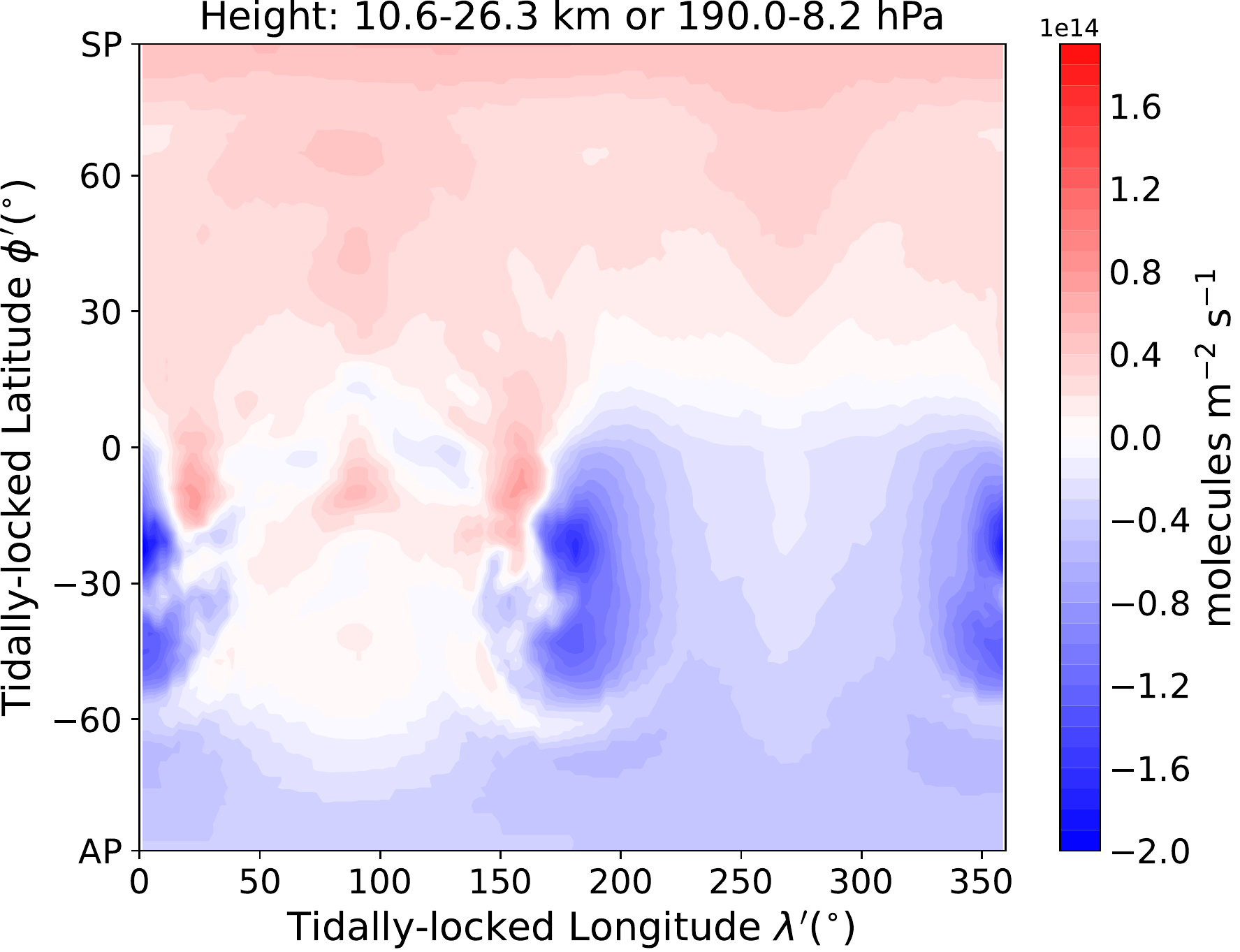}
\caption{Vertical flux of ozone (F$_{\rm{O_3}}$ in molecules m$^{-2}$ s$^{-1}$) between $P_{max}{=}190$~hPa and $P_{min}{=}8.2$~hPa. The predominantly downward exchange at the locations of the Rossby gyres illustrates how the enhanced ozone columns are driven by the downward motions that are part of the stratospheric dayside-to-nightside circulation.}
\label{fig:pcb_vert_o3flux_column}
\end{figure}

To further diagnose the nightside descent of ozone molecules indicated by the streamfunctions, we can define the vertical flux of ozone across pressure or altitude levels as:
\begin{equation}\label{eq:o3flux}
    F_{\rm{O_3}} = \int^{P_{min}}_{P_{max}} (w{\cdot}n_{\rm{O_3}}) dP,
\end{equation}
where $w$ is the vertical wind velocity (m~s$^{-1}$) and $n_{\rm{O_3}}$ the ozone number density in molecules~m$^{-3}$. Negative values correspond to downward transport and positive values to upward transport of ozone. The integration between pressure levels $P_{max}$ and $P_{min}$ is done to determine the total flux exchange between the stratosphere and troposphere. Using the streamfunctions in Figure~\ref{fig:streamfunctions} and the ozone distribution in Figure~\ref{fig:pcb_o3distrib}b, we determine that downward transport between ${\sim}200$ and 8~hPa drives the ozone accumulation. Figure~\ref{fig:pcb_vert_o3flux_column} shows the vertical flux of ozone, integrated over pressures between $190$ and 8~hPa. Generally, we find a relatively small but hemisphere-wide upward flux on the dayside. The nightside gyre locations stand out with a relatively strong downward flux. Hence, the ozone that was produced in the stratosphere will be transported downward into the troposphere at the gyre locations. Combining the streamfunctions, the tracer experiment and the vertical ozone flux, we find that the stratospheric overturning circulation provides a connection between the ozone production regions and the nightside gyres, leading to the accumulation of ozone in the latter. To the authors' knowledge, this is the first time this connection has been reported.

\subsection{Dynamical and chemical timescales}
In assessing the impact of atmospheric dynamics on chemical abundances, it is important to make a comparison between the timescales of processes that can control the ozone abundance. The dynamical lifetimes include the zonal ($\tau_{u}$), meridional ($\tau_{v}$), and vertical components ($\tau_{w}$), and are calculated following \citet{drummond_observable_2018}:
\begin{equation}
    \tau_{u} = \frac{L}{u} = \frac{2\pi R_p}{u},
\end{equation}
\begin{equation}
    \tau_{v} = \frac{L}{v} = \frac{\pi R_p}{v},
\end{equation}
\begin{equation}
    \tau_{w} = \frac{H}{w},
\end{equation}
with $L$ the relevant horizontal scale in terms of the planetary radius $R_\mathrm{p}$, and $H$ the vertical scale height. The zonal ($u$), meridional ($v$), and vertical ($w$) wind components are all in m/s. For the chemical lifetimes we use:
\begin{equation}
    \tau_{chem} = \frac{n_{\rm{O_3}}}{R_x},
\end{equation}
where $n_{\rm{O_3}}$ denotes the ozone number density (molecules~m$^{-3}$) and $R_x$ the loss of ozone (in molecules~m$^{-3}$~s$^{-1}$) due to reaction $x$. Specifically, we use the termination reaction of the Chapman mechanism \citep[][]{chapman_xxxv_1930, braam_lightning-induced_2022}:\\
\\
\ce{O$_3$ + O($^3$P) -> O$_2$ + O$_2$},  \hfill (R1) \\
\\
and the rate-limiting step of the dominant HO$_x$ catalytic cycle \citep[][]{yates_ozone_2020,braam_lightning-induced_2022}:\\
\\
\ce{HO$_2$ + O$_3$ -> OH + 2O$_2$}. \hfill (R2)\\
\\
A detailed overview of the chemical reactions can be found in \citet{braam_lightning-induced_2022}. We calculate the lifetimes for sets of gridpoints centred at four distinct locations in the ozone distribution (see Figure~\ref{fig:pcb_o3distrib}), and subsequently take the meridional and zonal mean. These locations cover the substellar point (10 latitudes ${\times}$ 8 longitudes = 80 grid points), the nightside jet ($10{\times}7{=}70$~points), and the two nightside gyres with $5{\times}7{=}35$~points each.

\begin{figure}
\includegraphics[width=\columnwidth]{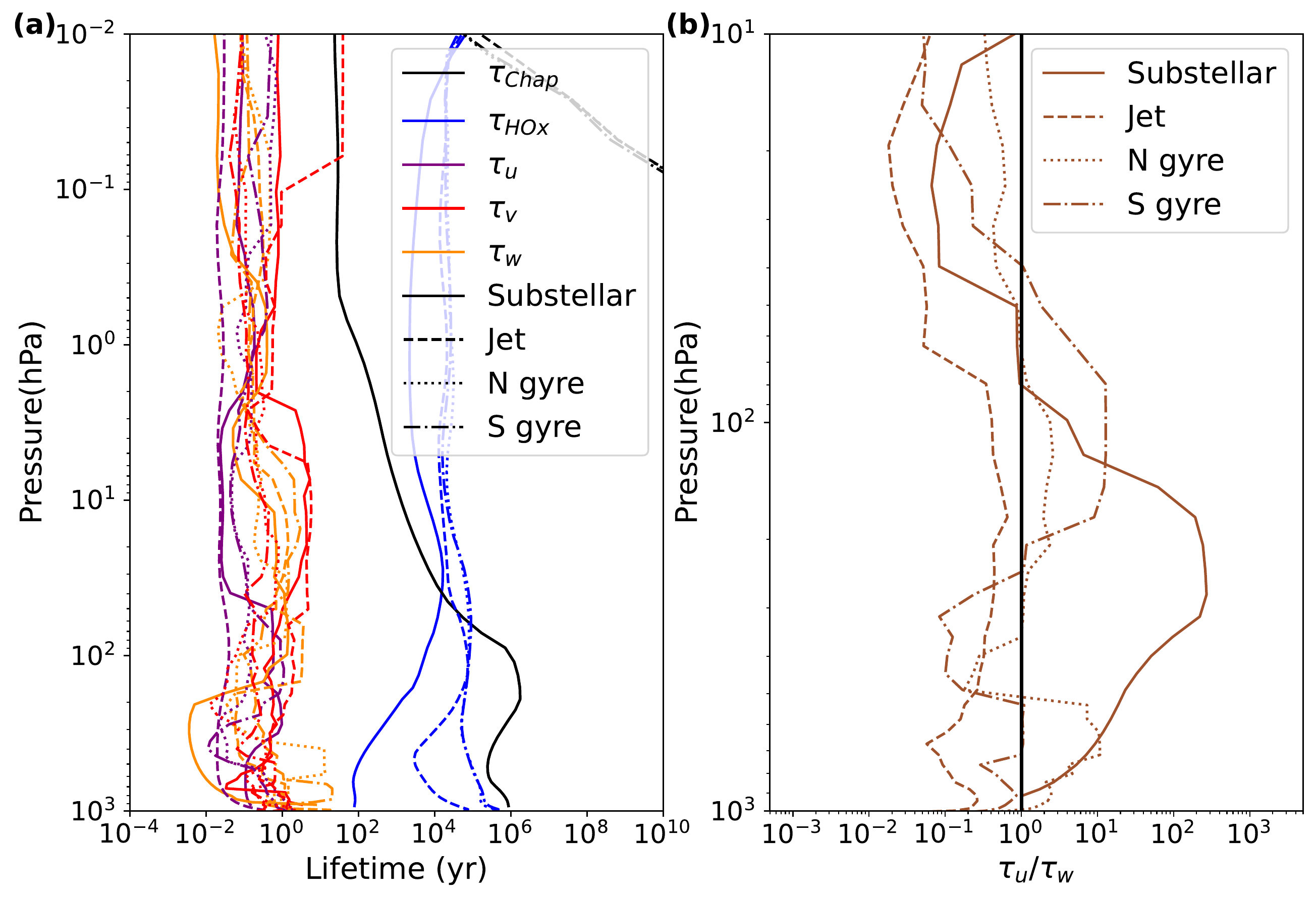}
\caption{Dynamical and chemical lifetimes over four locations in the atmosphere: the substellar point, two regions over the gyres and one region over the nightside jet. (a) $\tau_{u}$, $\tau_{v}$ and $\tau_{w}$ denote the dynamical lifetime versus zonal, meridional, and vertical transport, respectively. $\tau_{\mathrm{Chap}}$ and $\tau_{\mathrm{HOx}}$ show the chemical lifetimes of ozone versus loss by the Chapman termination reaction (R1) and the dominant HO$_{\rm{x}}$ catalytic cycle (R2), respectively. From the comparatively long chemical lifetimes we can deduce that dynamical processes control the chemical abundances. (b) The fraction of the zonal to vertical dynamical lifetimes in the lower stratosphere (between $100$~hPa and $10$~hPa) and troposphere (${<}100$~hPa), along with a vertical line indicating where they are equal ($\tau_{u}{/}\tau_{w}{=}1$). Vertical transport is the dominant process for $\tau_{u}{/}\tau_{w}{>}1$ and horizontal transport for $\tau_{u}{/}\tau_{w}{<}1$.}
\label{fig:pcb_lifetimes}
\end{figure}

Figure~\ref{fig:pcb_lifetimes} shows the different lifetimes at each of the four locations. From Figure~\ref{fig:pcb_lifetimes}a we conclude that the dynamical lifetimes are shorter than the chemical lifetimes at all four locations, indicating that dynamics can be an important driver of disequilibrium abundances in this pressure range. In Figure~\ref{fig:pcb_lifetimes}b we highlight the differences between $\tau_{u}$ and $\tau_{w}$, for the troposphere (${<}100$~hPa) and lower stratosphere (between $100$~hPa and $10$~hPa), by using the fraction $\tau_{u}{/}\tau_{w}$. Vertical transport is the dominant process for $\tau_{u}{/}\tau_{w}{>}1$ (right of the vertical line) and horizontal transport for $\tau_{u}{/}\tau_{w}{<}1$ (left of the vertical line). Around the substellar point (solid lines), we determine that vertical mixing dominates the troposphere ($\tau_{u}{/}\tau_{w}{>}1$) and that zonal mixing ($\tau_u$) starts to take over at P${>}80$~hPa. Above this pressure, chemical abundances at the substellar point can be spread out zonally towards the nightside, connecting with the ozone-producing region that is part of the overturning circulation from Section~\ref{subsec:circulation}. At the nightside location of the jet, $\tau_{u}{/}\tau_{w}{<}1$, and the zonal wind is capable of homogenising any vertically-driven disequilibrium. The circumnavigating jet then leads to the relatively thin ozone column for $70^{\circ}{<}\lambda'{<}110^{\circ}$ and $220^{\circ}{<}\lambda'{<}320^{\circ}$ in Figure~\ref{fig:pcb_o3distrib} (across all $\phi'$). At the locations of the nightside gyres, Figure~\ref{fig:pcb_lifetimes}b shows that $\tau_{u}$ and $\tau_{w}$ are intermittently the smallest, indicating that both vertical and zonal mixing can drive disequilibrium abundances. However, as mentioned in Section~\ref{subsec:circulation}, the edges of the gyres act as mixing barriers. Hence, the zonal transport leads to homogenisation within the gyres. Vertical mixing that is part of the overturning dayside-to-nightside circulation is dominant between ${\sim}$200 and 50~hPa at the gyre locations. This vertical mixing drives the observed disequilibrium abundances of tropospheric ozone at the gyre locations, and thus the maximum ozone columns in Figure~\ref{fig:pcb_o3distrib}a.

\section{Discussion}\label{sec:discussion}
In this section, we start by describing the driving mechanism for the overturning circulation. We then show its impact on other long-lived tracers and discuss relevant temporal variability in the atmospheres of synchronously rotating exoplanets. Lastly, we produce synthetic emission spectra to investigate the observational impact of circulation-driven ozone chemistry.

\subsection{Driving mechanism of the overturning circulation}
The tropospheric overturning circulation for moist, rocky exoplanets in a synchronous orbit is driven by the absorption of incoming stellar radiation and latent heat release on the dayside, and longwave radiative cooling on the nightside \citep[e.g.][]{showman_atmospheric_2013, boutle_exploring_2017}. \citet{wang_atmospheric_2022} study dry, rocky planets rotating synchronously around an M-dwarf star and find that the overturning circulation is indirectly driven by the stellar radiation, in the form of nightside cooling by CO$_2$. They find that an overturning circulation forms in a N$_2$-CO$_2$ atmosphere, but not in a pure N$_2$ atmosphere \citep[][]{wang_atmospheric_2022}. Prescribed CO$_2$ distributions from \citet{wang_atmospheric_2022} show that shortwave (SW) absorption on the planetary dayside only has a limited impact on the overturning circulation. CO$_2$ can cool an atmosphere when it is found in layers exhibiting a temperature inversion \citep[][]{wang_atmospheric_2022}. Enhanced infrared emission from increasing CO$_2$ levels cools the Earth's stratosphere \citep[][]{luther_temperature_1977, brasseur_stratospheric_1988, langematz_thermal_2003, shine_comparison_2003, fomichev_response_2007}. On synchronously rotating planets, this can induce a downward motion on the nightside that subsequently drives dayside-to-nightside overturning circulation.

\begin{figure}
\includegraphics[width=0.5\columnwidth]{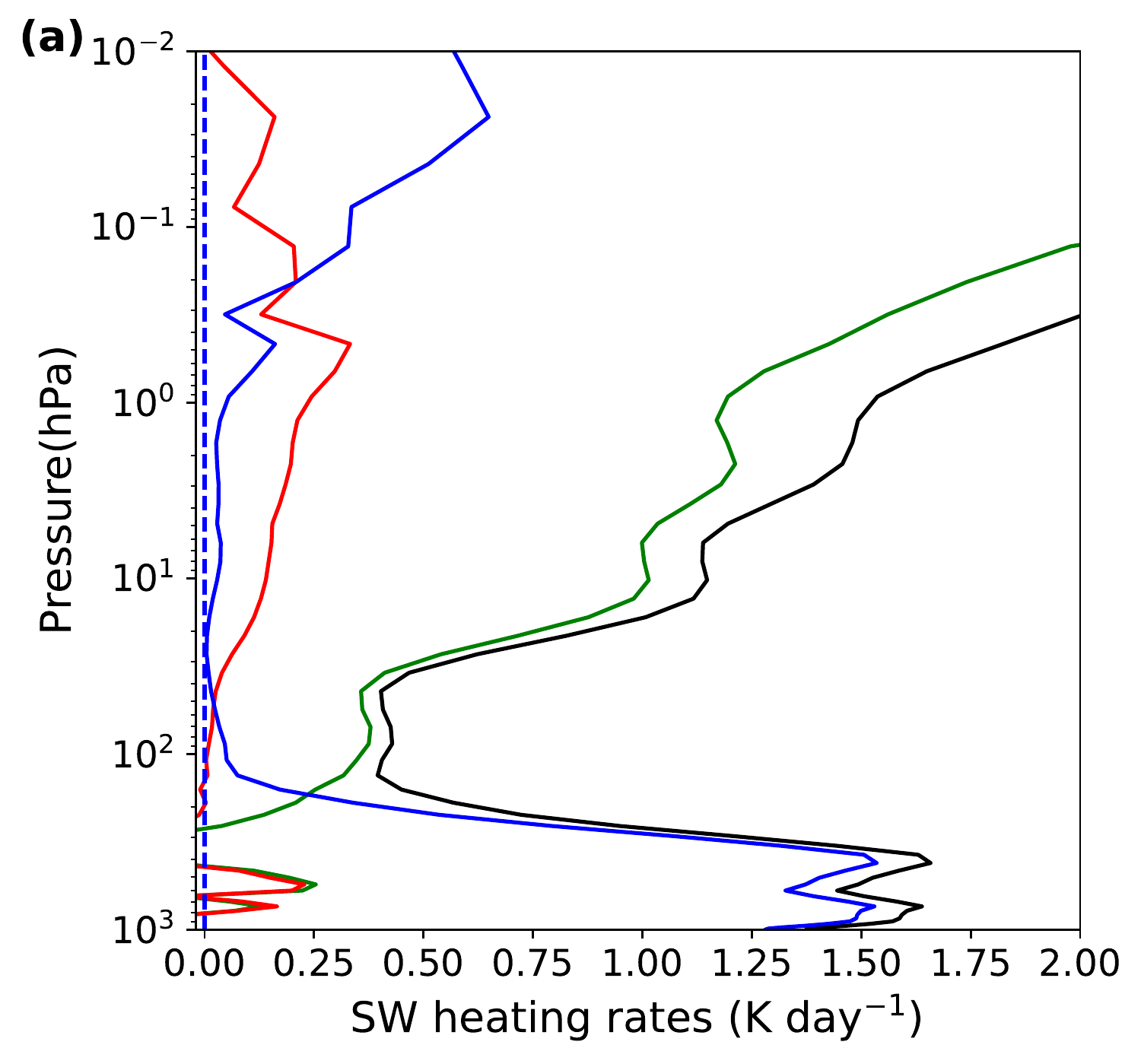}
\includegraphics[width=0.5\columnwidth]{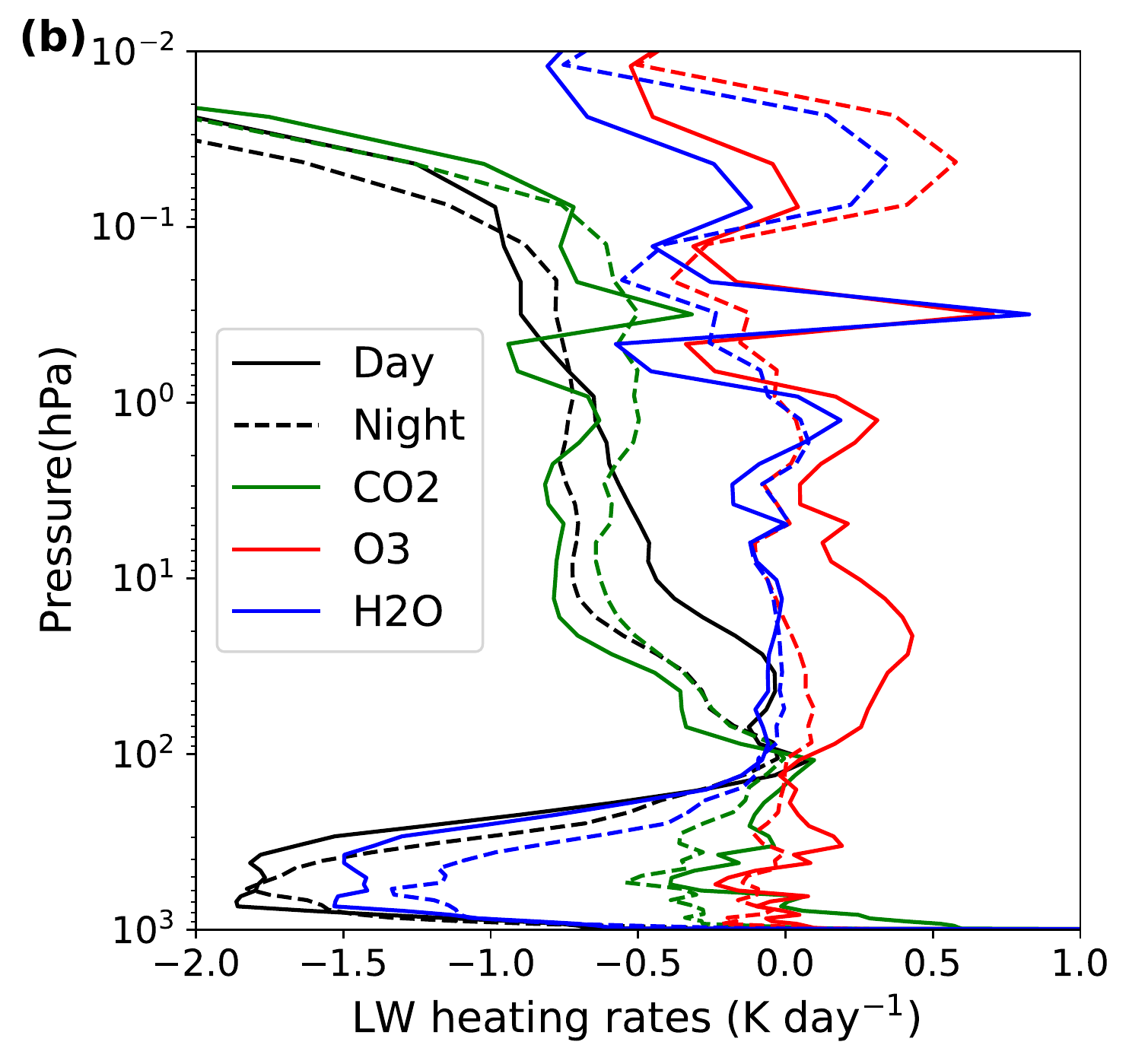}
\caption{Atmospheric heating (and cooling) rates, for (a) the SW radiation and (b) LW radiation. Solid lines show the hemispheric average over the dayside and dashed lines over the nightside. The coloured lines indicate the individual components contributing to the total heating rates in black, showing that CO$_2$ (green) becomes the dominant contributor to the dayside SW heating rates and that LW cooling is also mainly driven by CO$_2$.}
\label{fig:heatrates}
\end{figure}

Since we focus on the stratosphere, which is relatively dry even for a moist climate of a rocky exoplanet in a synchronous orbit, we can build upon these results in identifying the driving mechanism. The SW atmospheric heating rates in Figure~\ref{fig:heatrates}a show that CO$_2$ (the green line) acts as an important SW absorber on the dayside. The main absorber in the troposphere is H$_2$O, whereas CO$_2$ starts to become dominant above ${\sim}170$~hPa. In line with \citet{wang_atmospheric_2022}, we find that heating due to SW absorption by CO$_2$ plays a minor role in the troposphere. However, in the stratosphere CO$_2$ absorption can become important because peak emissions from M-dwarfs are emitted at near-infrared (NIR) wavelengths, relatively long as compared to other stars. CO$_2$ (and H$_2$O) have strong NIR absorption bands \citep[][]{selsis_habitable_2007, turbet_habitability_2016, lobo_terminator_2023}, which explains why CO$_2$ is the dominant absorbing species above ${\sim}170$~hPa, in contrast to ozone in the Earth's stratosphere. As expected, the total dayside heating rates (solid black line) greatly exceed the nightside values (dashed line), forming a direct driver for the overturning circulation. Additionally, Figure~\ref{fig:heatrates}b shows the longwave (LW) heating rates, with negative values indicating cooling of the atmosphere. The black lines show stronger LW cooling on the nightside as compared to the dayside. Again, CO$_2$ is mainly responsible for these cooling rates, due to its presence in temperature inversion layers at ${\sim}100$ and ${\sim}1$~hPa. This radiative cooling on the nightside drives a large-scale downwelling which, together with SW heating on the dayside, supports the stratospheric overturning circulation \citep[][]{edson_atmospheric_2011, koll_temperature_2016, wang_atmospheric_2022}, and can explain the ozone maxima at the locations of the nightside gyres. The atmospheric pressure within the gyre is relatively low, analogous to the eye of tropical cyclones \citep[][]{schubert_distribution_2007}. Such a pressure gradient naturally induces downward transport at the gyre locations. An important follow-up to this study is to investigate the ozone distribution for a variety of rotation states \citep[see e.g.][]{carone_stratosphere_2018, haqq-misra_demarcating_2018, chen_habitability_2019} in light of the circulation-driven chemistry proposed here.
 
\subsection{Long-lived atmospheric tracers}
The impact of the overturning circulation goes beyond the spatial distribution of ozone, as is also evident from the distribution of the age-of-air tracer as shown in Figure~\ref{fig:ageair_tllat}. Any tracer, gaseous or non-gaseous phase, can continue to circulate as long as its chemical lifetime is much longer than the dynamical timescales. Hence, the overturning circulation is relevant for any so-called long-lived atmospheric tracer. To illustrate this, we performed similar analyses using the species-weighted streamfunction as defined in Section~\ref{sec:metrics} on the distributions of nitric acid (HNO$_3$) and dinitrogen pentoxide (N$_2$O$_5$). Both of these species are signatures of lightning-induced chemistry in our simulations \citep[][]{braam_lightning-induced_2022}. They are non-radical species with relatively long chemical lifetimes, mainly in the form of photolysis and wet deposition (rainout). In the dayside troposphere, the lifetimes against wet deposition are ${\sim}10^{-2}{-}10^2$~yr, while higher up in the atmosphere the lifetimes against photolysis are ${\sim}10{-}10^2$~yr. On the nightside, these loss processes are absent and thus their chemical lifetimes approach infinity. We calculate $\Psi'_{\rm{HNO_3}}$ and $\Psi'_{\rm{N_2O_5}}$ similar to Equation~\ref{eq:mm_stream_tl_o3}, and calculate the mean of each of the species-weighted streamfunctions over the troposphere (${>}10^2$~hPa) and mid-to-lower stratosphere ($1{<}$P${<}10^2$~hPa). The results are shown in Table~\ref{tab:stream_fractions}.
\begin{table}
	\centering
	\caption{Species-weighted streamfunctions  $\Psi'_{\rm{x}}$ in kg~s$^{-1}$ averaged over pressure levels corresponding to the troposphere and stratosphere. Shown for ozone and lightning-induced chemistry in the form of HNO$_3$ and N$_2$O$_5$.}
	\label{tab:stream_fractions}
	\begin{tabular}{llll} 
		\hline \vspace*{0.1cm}
		   & $\Psi'_{\rm{O_3}}$ & $\Psi'_{\rm{HNO_3}}$ & $\Psi'_{\rm{N_2O_5}}$ \\
		\hline
		Troposphere & $9.70{\times}10^{5}$ & $9.47{\times}10^{-1}$ & $4.44{\times}10^{-2}$ \\
		Lower stratosphere & $6.09{\times}10^{5}$ & $6.07{\times}10^{-3}$ & $1.50{\times}10^{-3}$ \\
            \hline
  \end{tabular}
\end{table}
The circulation cells weighted by HNO$_3$ and N$_2$O$_5$ are strongest in the troposphere, at ${\sim}0.95$ and ${\sim}0.04$~kg~s$^{-1}$, respectively, because of the strong overturning circulation here (see Figure~\ref{fig:streamfunctions}b). The troposphere is also the region where lightning flashes are predicted to occur and thus where HNO$_3$, N$_2$O$_5$, and their precursors are produced \citep[][]{braam_lightning-induced_2022}. The factor $10^6$ and $10^7$ difference with the ozone-weighted streamfunction in Table~\ref{tab:stream_fractions} is a consequence of the much lower predicted abundances of HNO$_3$ and N$_2$O$_5$. Moving up to the stratosphere, we find that the ozone-weighted streamfunction is similar to the streamfunction in the troposphere, providing the connection to the nightside gyres. For HNO$_3$ and N$_2$O$_5$, the streamfunction is ${\sim}30$ and 150 times lower in the stratosphere, due to low levels of stratospheric HNO$_3$ and N$_2$O$_5$ with the absence of lightning-induced chemistry at those pressure levels. Because of the lack of stratospheric HNO$_3$ and N$_2$O$_5$, the overturning circulation will not be able to accumulate these species at the locations of the nightside gyres (as is evident in the spatial distribution in Figure~10 of \citealt{braam_lightning-induced_2022}). 

In the presence of stellar flares, \citet{ridgway_3d_2023} show that the gyres are depleted in ozone (see their Figure~12). This can also be explained by the stratospheric overturning circulation, since flare-induced chemistry will result in a large amount of nitric oxide (NO) and nitrogen dioxide (NO$_2$) (together known as the NO$_\mathrm{x}$ chemical family) at stratospheric levels \citep[][]{ridgway_3d_2023}. This NO$_\mathrm{x}$ can follow the stratospheric overturning circulation from the dayside to the nightside. Once on the nightside, it can be transported downward at the location of the gyres and locally deplete the ozone through the NO$_\mathrm{x}$ catalytic cycle \citep[][]{ridgway_3d_2023}, given that flares produce sufficient NO$_\mathrm{x}$.

The impact of the overturning circulation on the distribution of ozone has analogies with studies that simulate tracers in the atmospheres of synchronously rotating hot Jupiters. \citet{parmentier_3d_2013} identified dynamical mixing in hot Jupiter atmospheres as a process leading to cold trapping of condensible species on the planetary nightside. Their experiments involve gravitational settling as a source of these condensed particles, which leads to a gradient of tracer abundance, with fewer particles as we move up through the atmosphere. Upward mixing induced by the large-scale dynamics balances the settling of these particles, preventing the complete depletion of particles and inducing a strong spatial variation in the tracer abundances. The extent of the mechanism depends on the strength of frictional drag \citep[][]{komacek_vertical_2019}. The mechanism does not require convection but follows the large-scale atmospheric motions that are ultimately driven by the dayside-nightside heating contrast \citep[][]{parmentier_3d_2013}, as is the case for the circulation-driven ozone distribution discussed here. Another example of a long-lived tracer is photochemical haze, which is also expected to form at stratospheric altitudes \citep[e.g.][]{arney_pale_2017} and, for synchronously rotating exoplanets, only on the dayside of a planet \citep[e.g.][]{steinrueck_3d_2021}. \citet{steinrueck_3d_2021} show that the 3-D distribution of small photochemical hazes (${\leq}10$~nm) in hot Jupiter atmospheres is also driven by dynamical mixing. The highest tracer abundances are found above the production peak, indicating upwelling on the dayside. Then a divergent flow leads to transport towards the poles and the nightside. On the nightside, the haze particles are then advected downward and get trapped in the mid-latitude gyres \citep[][]{steinrueck_3d_2021}. These dynamically-induced asymmetries can produce distinctions between a planet's terminator regions, as shown for hot Jupiters \citep[][]{drummond_implications_2020, steinrueck_3d_2021, zamyatina_observability_2023}. Following up on the results presented here, we will investigate the potential terminator variability of the circulation-driven ozone distribution and its observability.

\subsection{Time variability}\label{sec:tempevol}
Besides spatial variability in tracer distributions, simulations of synchronously rotating exoplanets exhibit several modes of temporal variability. The formation of the Rossby gyres is due to the thermal forcing asymmetries \citep[][]{showman_matsuno-gill_2010, showman_equatorial_2011}{}{}. \citet{cohen_traveling_2023} show that these gyres oscillate over longitude $\lambda$, with the extent depending on the planet's rotation period and thus dynamical state. Planets with a slower rotation rate have longer oscillation periods, resulting in a 157.5-day oscillation for Proxima Centauri b, which was determined from the temporal evolution of the cloud cover \citep[][]{cohen_traveling_2023}.

\begin{figure}
\includegraphics[width=\columnwidth]{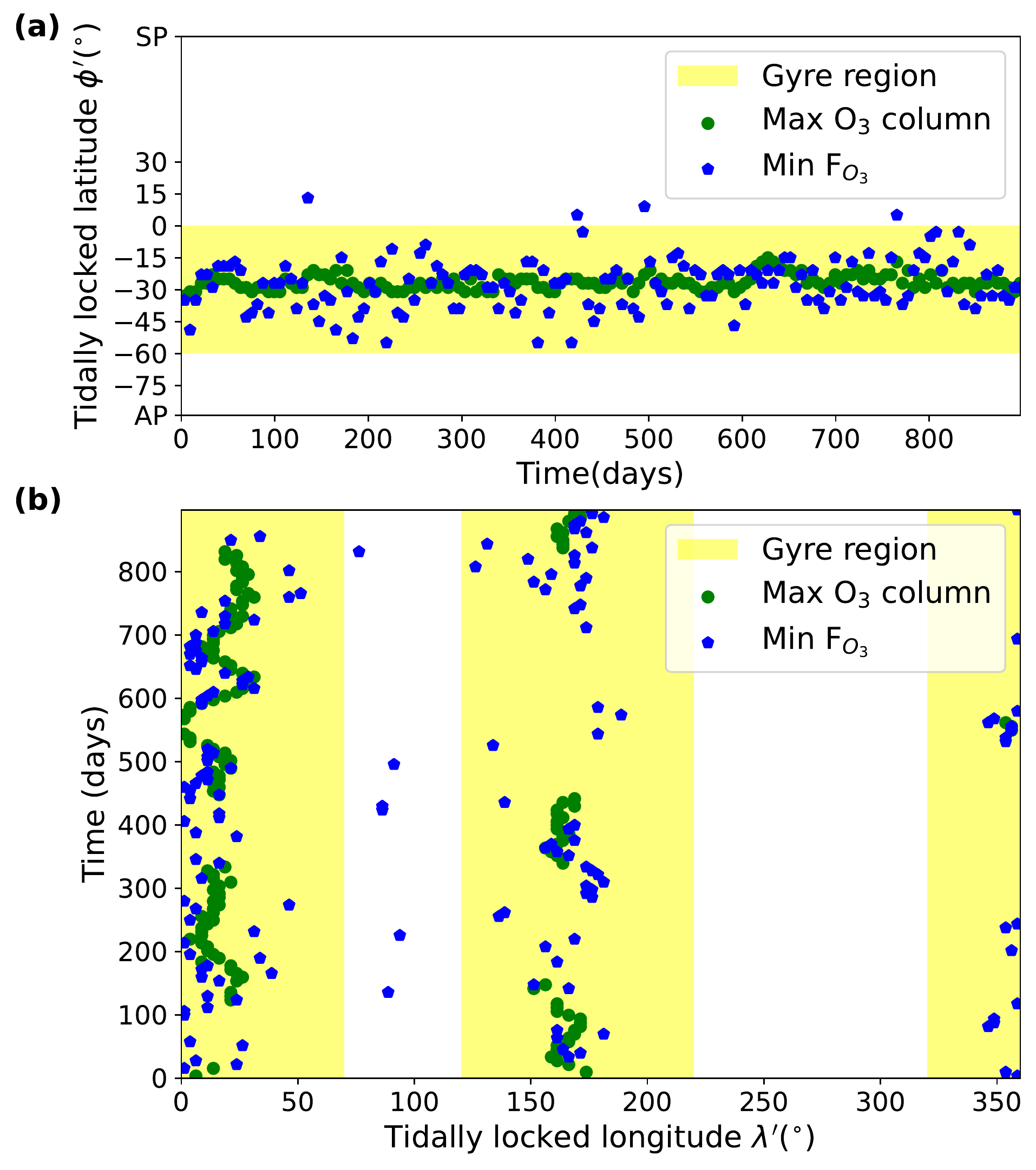}
\caption{Temporal evolution of the locations of extremes in the ozone column (see Figure~\ref{fig:pcb_o3distrib}) and vertical ozone flux $F_{\rm{O_3}}$ as defined in Equation~\ref{eq:o3flux}. We extract the tidally-locked latitude $\phi'$ and longitude $\lambda'$ corresponding to the maximum ozone column (shown as the green dots) and the minimum $F_{\rm{O_3}}$ (or the strongest downward flux, shown as the blue hexagons), to look for correlations between the two. Panels a and b show the temporal evolution of $\phi'$ and $\lambda'$, respectively, and the yellow rectangles indicate the gyre locations.}
\label{fig:ex_toc_o3f}
\end{figure}

Since the stellar spectra are constant in time and the planet rotates in a 1:1 resonant orbit without eccentricity and/or obliquity, such variability has to be produced by internal atmospheric variability. \citet{cohen_traveling_2023} show that feedback between cloud cover and the incoming stellar radiation can influence the dynamics and drive zonal movement by the gyres, leading to variations in humidity and cloud cover over time. The accumulation of ozone (Figure~\ref{fig:pcb_o3distrib}) depends on the gyres so we expect there also to be a corresponding variation in atmospheric ozone. To verify this, in Figure~\ref{fig:ex_toc_o3f} we track the temporal evolution of the tidally-locked coordinates corresponding to the maximum in the ozone layer and the minimum in the vertical flux of ozone (F$_{\rm{O_3}}$, thus corresponding to the strongest downward flux). Figure~\ref{fig:ex_toc_o3f}a shows $\phi'$ and Figure~\ref{fig:ex_toc_o3f}b $\lambda'$ corresponding to these extrema, and the approximate extents of the gyres are indicated in yellow. The locations of the maximum ozone column and minimum vertical flux are not perfectly aligned, because the maximum ozone column corresponds to a long-term mean location of the gyre and thus depends on vertical fluxes over an extended period of time. The minimum vertical flux represents a snapshot in time and is also impacted by the upward flux from the gyre (see the red regions in Figure~\ref{fig:pcb_vert_o3flux_column}). From Figure~\ref{fig:ex_toc_o3f}a, we determine that the maximum ozone column is generally found at $\phi'$ corresponding to the gyre locations, with a small meridional variation over time. The minimum F$_{\rm{O_3}}$ shows more variability in tidally-locked latitude, but the strongest downward flux is generally also located at the gyre locations. In Figure~\ref{fig:ex_toc_o3f}b, we see the variations in the tidally-locked longitude $\lambda'$ over time. The low-$\lambda'$ gyre typically hosts the maximum ozone column, but there are periods when the mid-$\lambda'$ gyre hosts the maximum in the ozone column. The variations in the minimum F$_{\rm{O_3}}$ broadly align with the maximum in the ozone column, following the gyre position that has the maximum ozone column at that time. The location of minimum F$_{\rm{O_3}}$ shows more variability due to its instantaneous nature.

\begin{figure}
\includegraphics[width=\columnwidth]{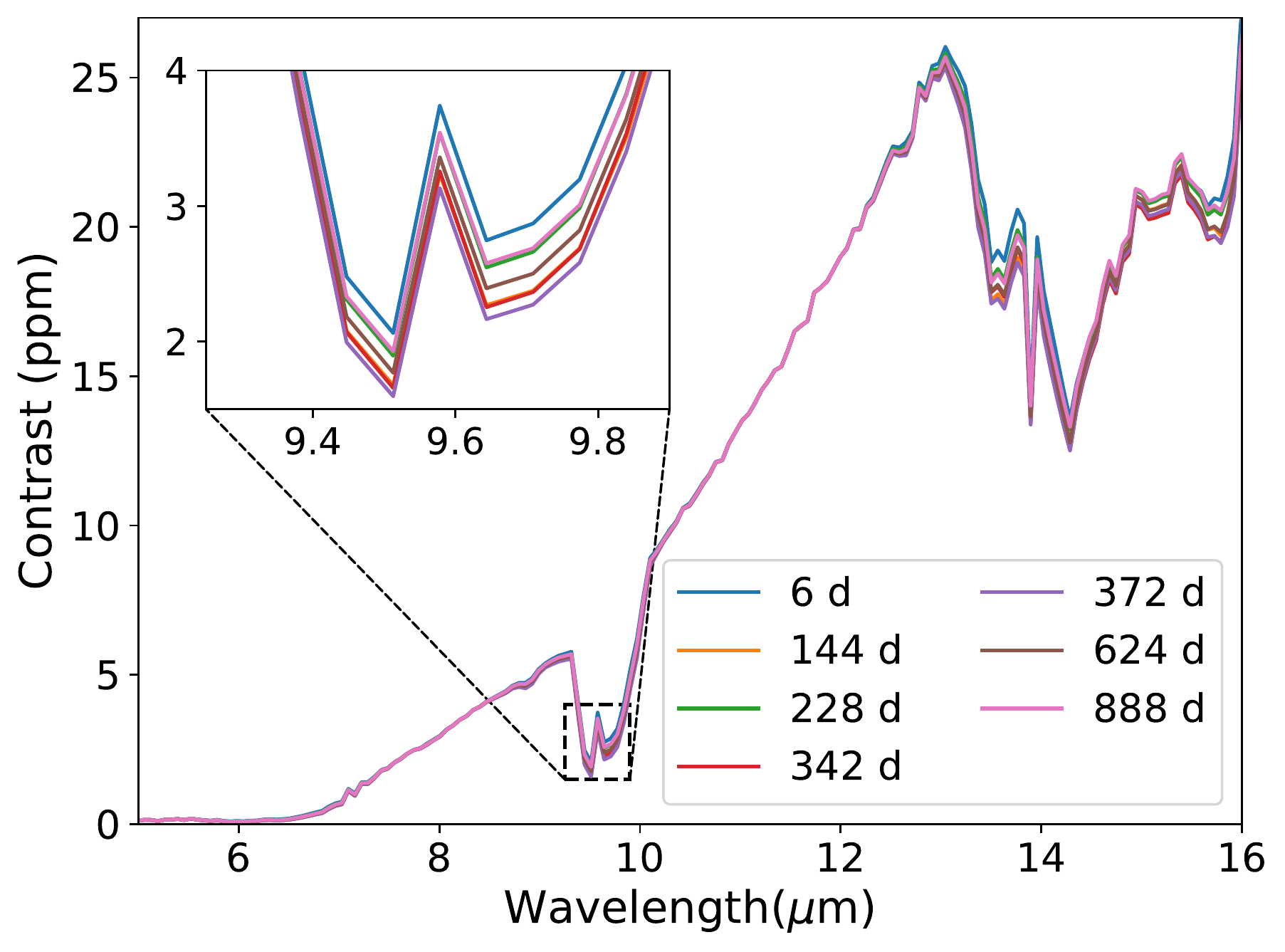}
\caption{Simulated emission spectra of Proxima Centauri b, for a range of 6-day intervals informed by extrema in the gyre locations (Figure~\ref{fig:ex_toc_o3f}). The legend shows the maximum day corresponding to each of the 6-day intervals. The inset region shows the region corresponding to the 9.6~$\mu$m ozone feature in greater detail.}
\label{fig:pcb_emission}
\end{figure}

We translate the temporal variability into simulated observables using the Planetary Spectrum Generator \citep[PSG:][]{villanueva_planetary_2018, villanueva_fundamentals_2022}. To simulate an emission spectrum that includes half the planetary dayside and half the nightside, we extract the atmospheric pressure and temperature and mixing ratios of relevant chemical species (N$_2$, O$_2$, CO$_2$, H$_2$O, O$_3$, N$_2$O, HNO$_3$ and N$_2$O$_5$) for these locations, take the zonal and meridional averages and compute radiative transfer with PSG. In Figure~\ref{fig:pcb_emission} we show the resulting planet-to-star contrast for the JWST-MIRI wavelength range, along with a zoom-in that focuses on the ozone 9.6 $\mu$m feature. Using extrema in the gyre positions over time from Figure~\ref{fig:ex_toc_o3f}, we simulate the emission spectra of Proxima Centauri b for different 6-day intervals and indicate the maximum day in the legend of Figure~\ref{fig:pcb_emission}. We find variations around the ozone features at 9.6~$\mu$m and between 14-16~$\mu$m that is due to absorption by CO$_2$, H$_2$O, and ozone. Hence, the region around 9.6 $\mu$m is the place to look for ozone variability. Focusing on the region around 9.6~$\mu$m shows that the maximum temporal variations are about 0.5~ppm. Spectroscopic characterisation of these absorption features to the level needed to identify these temporal variations is challenging, as detecting the features themselves would already require many days of co-added observations \citep[][]{kreidberg_prospects_2016}. However, the recent photometric observations of the thermal emission from TRAPPIST-1 b with JWST indicate the telescope's capacity to observe favourable terrestrial exoplanets \citep[][]{greene_thermal_2023}. Mission concepts such as the Large Interferometer For Exoplanets \citep[][]{quanz_large_2022} further utilise the mid-infrared in the characterisation of terrestrial exoplanets and will have to consider the impact of 3-D spatial and temporal variability in atmospheric dynamics and chemistry.

The hot Jupiter simulations of passive tracers by \citet{parmentier_3d_2013} also exhibit significant temporal variability. Oscillations in the equatorial jet and variations in the dayside-to-nightside flow produce large local variations, which could again impact the spectroscopic observations of the planets, both when conducting extended observations and when observing the same object at two different points in time.

Another mode of variability in the atmospheres of exoplanets in synchronous orbits around M-dwarfs is the Longitudinally Asymmetric Stratospheric wind Oscillation \citep[LASO;][]{cohen_longitudinally_2022}. Since this entails a stratospheric turnover of wind directions, it could be relevant for stratospheric ozone. Analysing ozone mixing ratios over time, we find variations in the ozone mixing ratios above ${\sim}30$~km (or ${\sim}3.5$~hPa) as a consequence of the LASO. However, these variations occur higher up in the atmosphere than the overturning circulation that feeds the gyres and thus do not affect the gyre abundances significantly. The variations are interesting from an observational perspective, which we plan to explore as part of an in-depth investigation of the observability of the circulation-driven ozone distribution.

\section{Conclusions}\label{sec:conclusion}
We use a 3-D CCM (UM-UKCA) to study the spatial structure of the ozone layer on an exoplanet rotating in a 1:1 spin-orbit resonance around an M-dwarf star, using the parameters corresponding to Proxima Centauri b. Our results are relevant for similar M-dwarf orbiting planets, specifically for slowly rotating planets with a strong overturning circulation and a single equatorial jet in the troposphere. We investigate the spatial variability in the ozone layer and specifically the accumulation in two nightside ozone maxima, in the form of maximum ozone columns at the locations of the permanent Rossby gyres. Our work builds upon previous studies that have shown that M-dwarf radiation supports the emergence of a global ozone layer. 

We show that stratospheric dayside-to-nightside circulation and downward motion over low-pressure nightside gyres can explain the spatial variability in ozone. The photochemistry required to initiate the Chapman mechanism of ozone formation is limited to the dayside hemisphere, with an absence of ozone production on the nightside. We find a connection between the ozone production regions on the dayside and the nightside hemisphere, using the transformation to the tidally-locked coordinate system. Meridional streamfunctions that we calculate from the divergent wind component illustrate the existence of a stratospheric dayside-to-nightside overturning circulation. This circulation consists of a single circulation cell characterized by upwelling motion in the ozone production regions, followed by stratospheric dayside-to-nightside transport and downwelling motions at the locations of the nightside gyres. The downwelling motion produces a flux of ozone from the stratosphere into the troposphere, leading to well-defined maxima in the ozone distribution. The circulation-driven ozone chemistry impacts spectroscopic observations, although the impact of temporal variability is limited to sub-ppm levels in emission spectra.

By investigating the impact of the stratospheric overturning circulation on lightning-induced chemical species (also limited to dayside production, but solely in the troposphere), we can explain why these species do not show a similar accumulation in the nightside gyres. The stratospheric overturning circulation also affects other tracer species, including gaseous chemical tracers and particulate components of photochemical haze, with the only requirement that the dynamical lifetimes are sufficiently short compared to chemical timescales.

We identify hemispheric contrasts in atmospheric heating and cooling rates as the driver for the overturning circulation. Dayside heating can directly drive the overturning circulation, and nightside cooling provides an indirect component by inducing local downward motion. The relatively low atmospheric pressure over the nightside gyres further induces downward motion here. Since the stratosphere is relatively dry, CO$_2$ absorption is the main contributor to these heating and cooling rates. Ozone absorption also contributes to the rates, but its contribution is weaker than CO$_2$ since M-dwarf fluxes peak close to absorption bands of CO$_2$. 

For the first time, we find a connection between the ozone-producing dayside of synchronously rotating planets and the simulated ozone maxima on the nightside, covering hemispheric scales and multiple vertical levels in the stratosphere and troposphere. The role of the stratospheric dayside-to-nightside circulation in driving the ozone distribution around the planet illustrates the necessity of 3-D model to capture atmospheric processes correctly. Any robust interpretation of spectroscopic observations will need to understand the spatial and temporal variability of chemical species due to such circulation-driven chemistry.

\section*{Acknowledgements}
We are very grateful to Denis Sergeev for his contribution to the coordinate transformations and valuable feedback on the manuscript. MB kindly thanks Ludmila Carone for discussing circulation regimes on synchronously rotating exoplanets.

MB, PIP and LD are part of the CHAMELEON MC ITN EJD which received funding from the European Union’s Horizon 2020 research and innovation programme under the Marie Sklodowska-Curie grant agreement no. 860470. PIP acknowledges funding from the STFC consolidator grant \#ST/V000594/1. LD acknowledges support from the KU Leuven IDN grant IDN/19/028 and from the FWO research grant G086217N. MC acknowledges the funding and support provided by the Edinburgh Earth, Ecology, and Environmental Doctoral Training Partnership and the Natural Environment Research Council [grant No. NE/S007407/1]. NM was supported by a UKRI Future Leaders Fellowship [grant number MR/T040866/1],  a Science and Technology Facilities Council Consolidated Grant [ST/R000395/1] and  the Leverhulme Trust through a research project grant [RPG-2020-82]. 

We gratefully acknowledge the use of the MONSooN2 system, a collaborative facility supplied under the Joint Weather and Climate Research Programme, a strategic partnership between the Met Office and the Natural Environment Research Council. Our research was performed as part of the project space ‘Using UKCA to investigate atmospheric composition on extra-solar planets (ExoChem)'. For the purpose of open access, the authors have applied a Creative Commons Attribution (CC BY) licence to any Author Accepted Manuscript version arising from this submission.

\section*{Data Availability}
All the CCM data was generated using the Met Office Unified Model and UK Chemistry and Aerosol model (\href{https://www.ukca.ac.uk/}{https://www.ukca.ac.uk/}), which are available for use under licence; see \href{http://www.metoffice.gov.uk/research/modelling-systems/unified-model}{http://www.metoffice.gov.uk/research/modelling-systems/unified-model}. The data underlying this article will be shared on reasonable request to the corresponding author, mainly motivated by the size of the data.

We used the iris \citep[][]{met_office_iris_2022} and aeolus \citep[][]{sergeev_aeolus_2022} python packages for the post-processing of model output. Scripts to process and visualize the data are available on github: \href{https://github.com/marrickb/o3circ\_code}{https://github.com/marrickb/o3circ\_code}.



\bibliographystyle{mnras}
\bibliography{o3circ.bbl} 




\bsp	
\label{lastpage}
\end{document}